\documentclass[article,5p,times,twocolumn]{elsarticle}           
\usepackage{amsmath}
\usepackage{amsfonts}
\usepackage{amssymb}
\usepackage{graphicx}
\usepackage{dcolumn}
\usepackage{txfonts}
\usepackage{float}
\usepackage{lineno}
\usepackage{bm}
\usepackage[ulem=normalem]{changes}
\usepackage{ulem}
\usepackage{todonotes} 
\usepackage{color}
\usepackage{booktabs}
\usepackage{upgreek}
\usepackage[ulem=normalem]{changes}
\usepackage{lineno,hyperref}
\usepackage[utf8]{inputenc}
\modulolinenumbers[5]

\def \micro#1{\upmu#1}  
\DeclareUnicodeCharacter{0301}{\'{e}}

\begin{document}
\begin{frontmatter}
\title{The MUGAST-AGATA-VAMOS campaign : set-up and performances}
\fntext[fn1]{\emph{Present address:} Normandie Univ, ENSICAEN, UNICAEN, CNRS/IN2P3, LPC Caen, 14000 Caen, France} 
\fntext[fn2]{\emph{Present address:} TU Darmstadt, Schlossgartenstra\ss e 9, D-64289 Darmstadt, Germany } 
\cortext[mycor1]{Corresponding author}
\author [ijc]{M. Assi\'e\corref{mycor1}}
\ead{marlene.assie@ijclab.in2p3.fr}
\author [ganil] {E. Cl\'ement}
\author [ganil] {A. Lemasson}
\author [ganil] {D. Ramos}
\author [infn] {A. Raggio}
\author[infn,unife] {I. Zanon}
\author [ijc] {F. Galtarossa}
\author[lpc] {C. Lenain}
\author[unipd] {J. Casal}
\author [ijc]{ F. Flavigny\fnref{fn1}}
\author [lpc] {A. Matta}
\author [unipd, infnpd] {D. Mengoni}
\author [ijc] {D. Beaumel}
\author [ijc] {Y. Blumenfeld}
\author [ro]{R. Borcea}
\author[infn,unipd]{D. Brugnara}
\author [surrey] {W. Catford}
\author [ganil]{F. de Oliveira}
\author [ijc]{N. De S\'er\'eville}
\author [iphc]{F. Didierjean}
\author [york]{C. Aa. Diget}
\author [ipnl] {J. Dudouet}
\author [sj]{B. Fern\'andez-Dom\'inguez}
\author [ganil]{C. Foug\`eres}
\author[ganil]{G. Fr\'emont}
\author[ganil, ijc]{V. Girard-Alcindor\fnref{fn2}}
\author[ganil]{A. Giret}
\author[infn]{A. Goasduff}
\author[infn]{A. Gottardo}
\author[ganil]{J. Goupil}
\author[ijc]{F. Hammache}
\author[tu] {P.R. John}
\author[ijc]{A. Korichi}
\author[ijc]{L. Lalanne}
\author[ganil]{S. Leblond}
\author [ganil] {A. Lefevre}
\author [ganil] {F. Legruel}
\author[ganil] {L. M\'enager}
\author[milano]{B. Million}
\author [ganil] {C. Nicolle}
\author [lpc] {F. Noury}
\author[ijc] {E. Rauly}
\author[iphc]{K. Rezynkina}
\author[ijc] {E. Rindel}
\author[york]{J.S. Rojo}
\author[cea] {M. Siciliano}
\author[ro]{M. Stanoiu}
\author[ijc]{I. Stefan}
\author[ijc] {L. Vatrinet}

\address[ijc]{Universit\'e Paris-Saclay, CNRS/IN2P3, IJCLab, 91405 Orsay, France}
\address[ganil]{ Grand Acc\'el\'erateur National d’Ions Lourds (GANIL), CEA/DRF-CNRS/IN2P3, Bvd Henri Becquerel, 14076 Caen, France }
\address[infn]{INFN, Laboratori Nazionali di Legnaro, viale dell'Universit\`a, 2-35020 Legnaro, Italy}
\address[unife]{Dipartimento di Fisica e Scienze della Terra, Universit\`a di Ferrara, via G. Saragat, 1-44121 Ferrara, Italy}
\address[lpc]{Normandie Univ, ENSICAEN, UNICAEN, CNRS/IN2P3, LPC Caen, 14000 Caen, France}
\address[unipd]{Dipartimento di Fisica, Universit\`a di Padova, via F. Marzolo, 8-35131 Padova, Italy}
\address[infnpd]{INFN Sezione di Padova, via F. Marzolo, 8 - 35131 Padova, Italy.}
\address[ro]{Horia Hulubei National Institute of Physics and Nuclear Engineering, Magurele, Romania}
\address[surrey] {Department of Physics, University of Surrey, Guildford GU2 5XH, United Kingdom}
\address[iphc]{Universit\'e de Strasbourg, IPHC, 23 rue du Loess, 67037 Strasbourg, France}
\address[york]{Department of Physics, University of York, York YO10 5DD, UK}
\address[ipnl] {Universit\'e de Lyon, Univ. Claude Bernard Lyon 1, CNRS/IN2P3, IP2I Lyon, F‐69622, Villeurbanne, France}
\address[sj]{IGFAE and Dpt. de F\'{\i}sica de Part\'{\i}culas, Univ. of Santiago de Compostela, E-15758, Santiago de Compostela, Spain}
\address[tu]{TU Darmstadt, Schlossgartenstra\ss e 9, D-64289 Darmstadt, Germany}
\address[milano]{INFN Sezione di Milano, IT-20133 Milano, Italy}
\address[cea]{CEA, Centre de Saclay, IRFU, F-91191 Gif-sur-Yvette, France}
%

\begin{abstract}
The MUGAST-AGATA-VAMOS set-up at GANIL combines the MUGAST highly-segmented silicon array with the state-of-the-art AGATA array and the large acceptance VAMOS spectrometer. The mechanical and electronics integration copes with the constraints of maximum efficiency for each device, in particular $\gamma$-ray transparency for the silicon array. This complete set-up offers a unique opportunity to perform exclusive measurements of direct reactions with the radioactive beams from the SPIRAL1 facility. The performance of the set-up is described through its commissioning and two examples of transfer reactions measured during the campaign. High accuracy spectroscopy of the nuclei of interest, including cross-sections and angular distributions, is achieved through the triple-coincidence measurement. In addition, the correction from Doppler effect of the $\gamma$-ray energies is improved by the detection of the light particles and the use of two-body kinematics and a full rejection of the background contributions is obtained through the identification of heavy residues. Moreover, the system can handle high intensity beams (up to 10$^8$ pps). The particle identification based on the measurement of the time-of-flight between MUGAST and VAMOS and the reconstruction of the trajectories is investigated.
\end{abstract}

\end{frontmatter}


\section{Introduction}
\label{intro}

Direct reactions are one of the keys to our understanding of nuclear structure and nuclear astrophysics. Their reaction mechanism involves few degrees of freedom and allows precise theoretical calculations such that they can supply crucial nuclear structure information. Transfer reactions, in particular, are an efficient tool to study nuclear structure effects such as shell evolution, pairing and clustering. Astrophysically important cross-sections can also be inferred through these measurements. For decades light ion beams were used to probe the heavier target nuclei of interest in direct kinematics. The inception of radioactive beams allowed the extension of direct reaction studies to the broad domain of unstable nuclei. In this case, the reactions are performed in inverse kinematics where the radioactive beam impinges on the light probe.
The excitation energy and scattering angle are generally determined through two-body kinematics from the kinetic energy and recoiling angle of the light particle. This technique has been employed by the first generation of Silicon-strip arrays such as MUST \cite{Blu99}, Orruba \cite{Bar13}... The straggling of the light charged particle in the target and the fact that dE/dE$^*$ (where E is the kinetic energy of the recoiling particle and E$^*$ the excitation energy) is smaller than 1 for pick-up reactions (kinematic compression) limits the excitation energy resolution attainable. Therefore, a simple particle measurement is only useful in cases when the level spacing is large, such as light or closed-shell nuclei. For more general cases, coupling of the particle detectors with a $\gamma$-ray array becomes crucial for precise spectroscopy and moreover brings additional information such as decay modes of the populated states.
%
A further combination with a zero-degree spectrometer provides event-by-event identification of the reaction products, separation between the beam and contaminants as well as rejection of the background contributions of fusion-evaporation residues from target-like products for example.

Measuring direct reactions in inverse kinematics and coupling with state-of-the-art $\gamma$-detector arrays requires the development of powerful and sophisticated devices. In particular, the detection of the light ejectile requires high accuracy both in angle and energy combined with compacity and as much as possible transparency to $\gamma$-rays. Silicon technology offers a combination of good energy resolution together with high granularity. TIARA \cite{tiara}, T-REX \cite{trex}, ShARC \cite{sharc} and GODDESS \cite{goddess} arrays have been built in this spirit with maximal solid-angle coverage and integration into $\gamma$-detector arrays but they offer limited particle identification. On the other hand, the MUST2 array \cite{must2} provides particle identification by $\Delta$E-E and time-of-flight techniques and can be arranged in a forward and backward four-telescope configuration to be coupled with 4 EXOGAM clovers \cite{exogam}  at 90 degrees \cite{Bur14,Gir17,Per20,Lec21}. Its poor transparency to $\gamma$-rays is due to its integrated electronics and cooling blocks which stand right behind the detectors and prevents coupling them with state-of-art HPGe $\gamma$-ray detectors such as AGATA \cite{agava,agata2,agata3}. 

The present paper focuses on the MUGAST-AGATA-VAMOS campaign at GANIL \cite{ganil} using the post-accelerated radioactive beams produced by the SPIRAL1 facility \cite{Spiral1}. The combination of the MUGAST Silicon array with the state-of-the-art AGATA array and the large acceptance VAMOS spectrometer \cite{Rej11} gives a unique opportunity to perform exclusive measurements with radioactive beams. The triple coincidence provides both the particle and $\gamma$-ray spectroscopy of the nucleus of interest, the entry point for the $\gamma$ decay through particle measurement, excellent background rejection, precise monitoring of the target thickness and Doppler correction from the two-body kinematics of the reaction. 

These features will be illustrated using three distinct data sets acquired with the MUGAST-AGATA-VAMOS setup: \emph{(i)} the commissioning data, where the well-studied $^{16}$O(d,p$\gamma$) transfer reaction was measured and provides insight into the excitation energy resolution and on the extraction of the angular distributions and, consequently, of the spectroscopic factors,  \emph{(ii)} the $^{19}$O(d,p$\gamma$) data set, which illustrates the energy resolution with a thin target and the use of two-body kinematics to perform event-by-event Doppler correction of the $\gamma$-rays energies, \emph{(iii)} the $^{15}$O($^{7}$Li,t$\gamma$) data set, demonstrating the background rejection obtained using triple coincidences and the possibility to run in high intensity mode. 

\begin{figure}[h!]
\begin{center}
\includegraphics[width=0.4\textwidth]{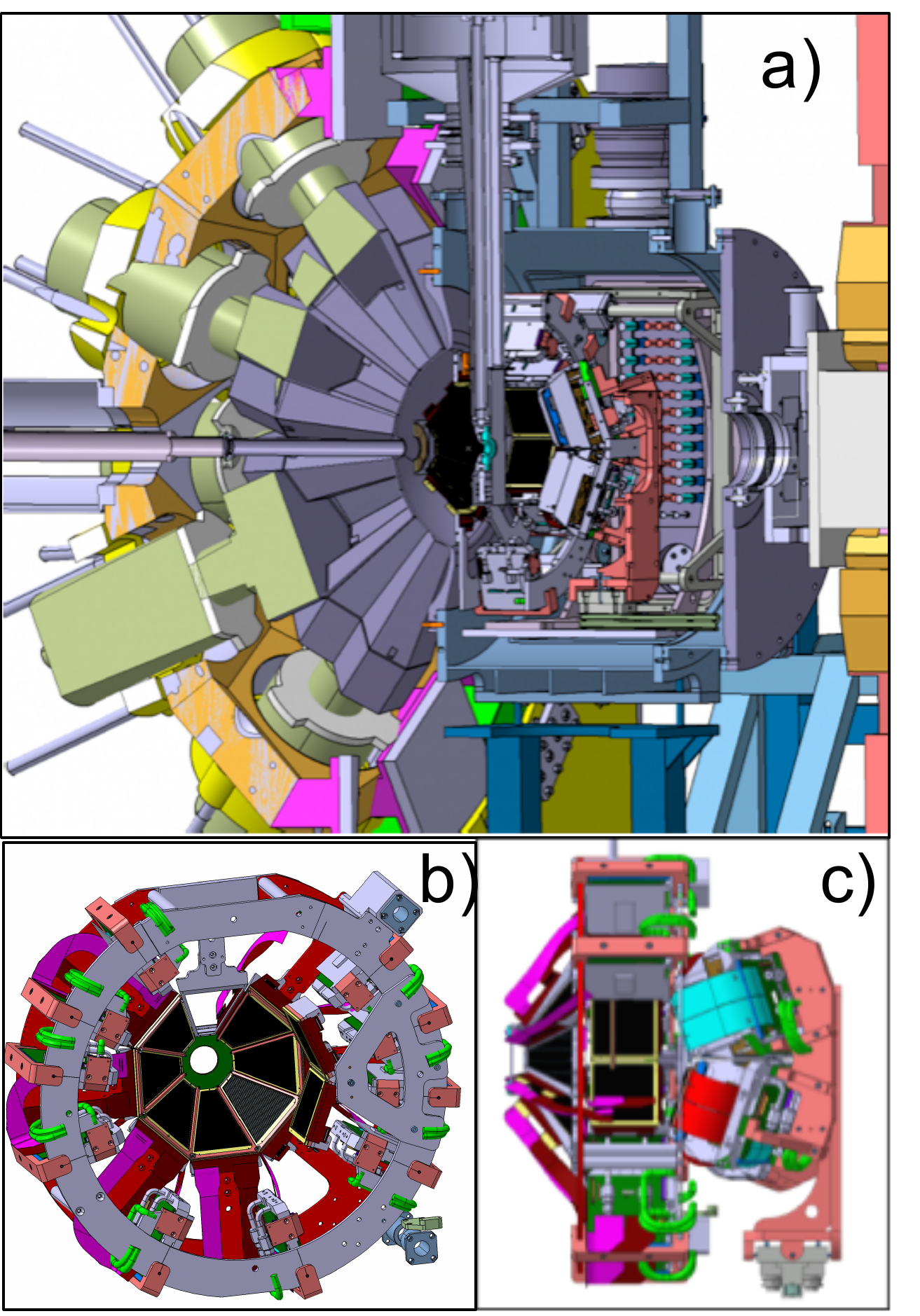}
\includegraphics[width=0.35\textwidth]{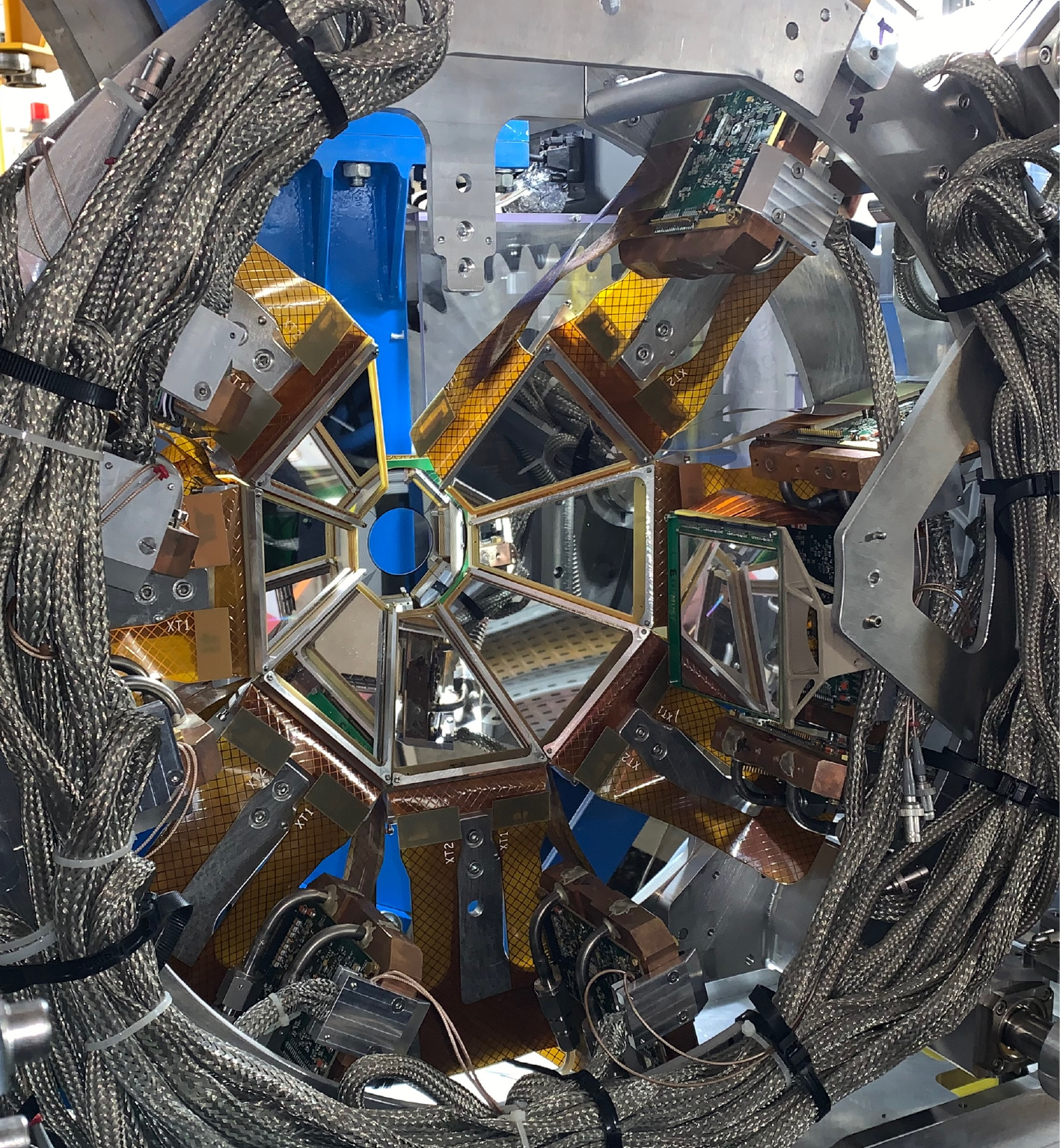}
\caption{ (Top) 3D view of the MUGAST set-up: a) global view with the beam coming from the left, AGATA on the left and the VAMOS spectrometer on the right; b) backward ring; c) full array. (Bottom) Photograph of the backward detectors with their electronics. \label{design}}
\end{center}
\end{figure}

\section {Implementation of the MUGAST-AGATA-VAMOS set-up}

The MUGAST-AGATA-VAMOS campaign was performed at GANIL using the radioactive SPIRAL1 beams post-accelerated by the CIME cyclotron up to 12A\,MeV. Beam tracking devices, CATS \cite{cats}, were available on the line for beam diagnostics and also time-of-flight measurement and beam rate monitoring. 
The beam was impinging either on a solid target (such as CD$_2$ or LiF) or a cryogenic $^3$He target, HeCTOr \cite{Hector}. The low beam energy is well suited for stripping reaction measurements, for example (d,p$\gamma$), where the light ejectile is emitted backward. The MUGAST array surrounded the target to provide energy and angle of the light particles together with their identification. The AGATA array surrounded MUGAST in the backward hemisphere and featured high resolution with line shape information and high efficiency. Finally, the VAMOS large acceptance spectrometer was used as a separator to provide identification of the heavy residue, rejection of the parasitic reactions and target thickness monitoring. Each element of the set-up is detailed in the following subsections.

\subsection{The MUGAST array}
The MUGAST array is composed of 12 double-sided Silicon strip detectors (DSSD) (see Fig.\,\ref{design}) which are arranged in order to cope with the mechanical constraints of the AGATA array and to fit into a sphere of 18\,cm diameter. The transparency to $\gamma$-rays in the backward direction is ensured by deporting the front-end electronics to a circular mechanical support (see Fig.\,\ref{design}b)) located at 90 degrees.

In the backward direction, 5 (in 2019) and then 7 (starting from 2020) trapezoidal nTD type DSSDs and a float zone annular DSSD are assembled as shown in Fig.\,\ref{design} to cover about 50\% (70\% with 7 trapezoidal detectors) of the backward hemisphere.
The trapezoidal detectors are 500 $\micro$m-thick, reverse mounted and placed at 13\,cm from the target. Their custom packaging from Micron Semiconductors Ltd \cite{micron} features very narrow frames with two kapton cables bent at 80\,degrees. Each side is divided into 128 strips with a pitch of 710\,$\micro$m and 760\,$\micro$m on the front and rear side, respectively.
The annular detector is a customized 500\,$\micro$m-thick Si DSSD (S1 type) from Micron Semiconductor Ltd. It is divided into 4 quadrants each comprising 16 rings on the junction side and 4 sectors in the ohmic side. The inner diameter of the active zone is 48\,mm and the external one 96\,mm. The angular resolution of the backward array due to the geometry is better than 0.4\,degrees.

\begin{figure}[h!]
\begin{center}
\includegraphics[width=0.45\textwidth]{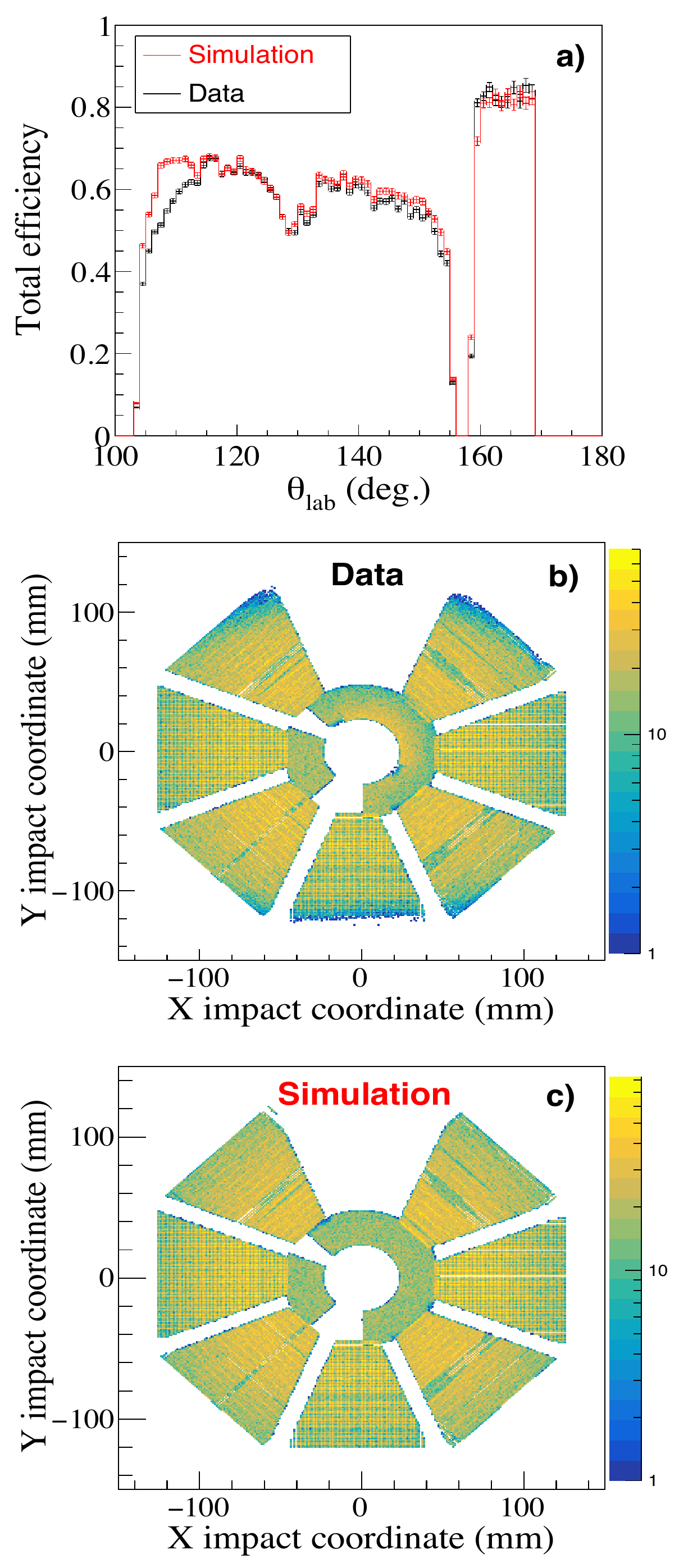}
\caption{a) Efficiency of MUGAST obtained through alpha source calibration for the configuration with 7 trapezoidal detectors in 2020. The red and black histograms represent simulation and data, respectively. The data is normalized to the simulation through the ratio of total number of counts in the annular detector. b) 2D reconstruction of the position of the alpha particles in the detectors. c) Same picture for simulated data.} \label{efficiency}
\end{center}
\end{figure}

 Fig.\,\ref{efficiency}a) shows the comparison between the measured (with an $\alpha$ source) and simulated total efficiencies of the backward detectors as a function of the position angle. The simulation is performed with the \emph{nptool} package \cite{nptool} in a realistic way i.e. taking into account the detector geometry, the dead layers and the missing strips during the experiment. The averaged efficiency is about 70\% for the trapezoidal ring which is consistent with the geometrical coverage. For the annular detector, the efficiency is close to 80\% as there were 3 missing sectors during the experiment (see Fig.\,\ref{efficiency}). There is a slight overestimation of the efficiency in the simulations around 110 degrees. This effect can be seen in Fig.\,\ref{efficiency}b) and c) where the impact matrix in the backward detectors is displayed. In the data, there are missing counts at the top and the bottom that are present in the simulated data. This loss of efficiency is attributed to the target holder frame, casting a shadow on the detectors.
 
 To complete the set-up, 4 MUST2 telescopes \cite{must2} cover the forward angles from 8 to 50\,degrees at 18\,cm from the target. They consist of a 300 $\micro$m-thick square DSSD backed by 16 CsI crystals read by photodiodes. The DSSDs are highly segmented with 128 strips on each side. Around 90\,degrees, a single square DSSD of 300\,$\micro$m thickness with 128 strips on each side is used to measure the elastic scattering of the beam on target. 

\begin{figure}[h!]
\begin{center}
\includegraphics[width=0.3\textwidth]{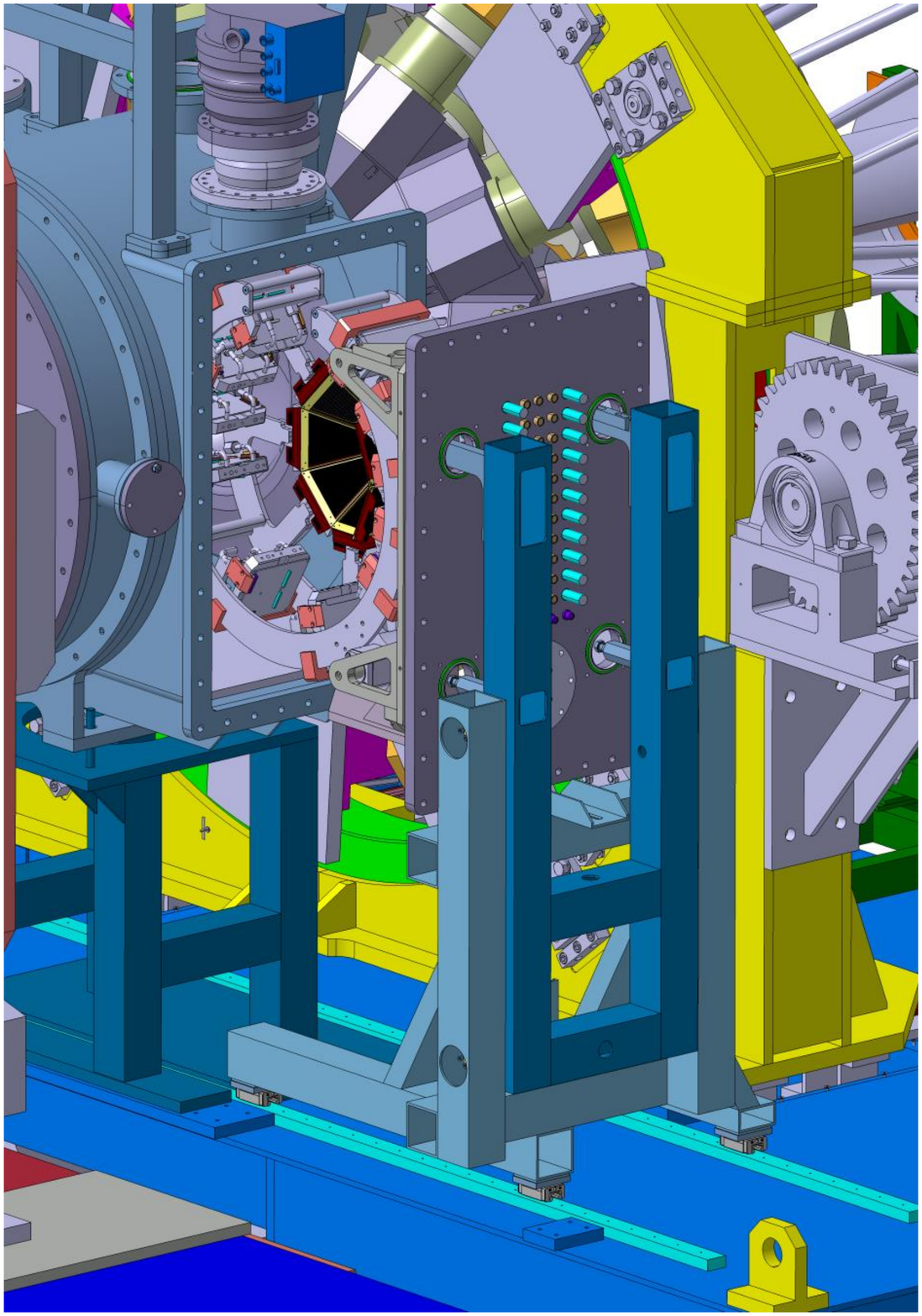}
\caption{ 3D view of the MUGAST reaction chamber and its mechanics.\label{meca}}
\end{center}
\end{figure}

The reaction chamber is designed such that the MUGAST array fits into the AGATA array and preserves the large angular acceptance of VAMOS. The backward removable part of the reaction chamber is an aluminium spherical shell of 18\,cm outer radius and a 3\,mm thickness. The central part of the reaction chamber is made of stainless steel and opens with two large doors, one on each side. A system of guide rails facilitates moving of the MUST2 detectors and the backward ring upstream by 20\,cm such that a moving cart can extract the backward ring out of the chamber and give full access to the detectors and their associated electronics (see Fig.\,\ref{meca}). The reaction chamber can also host the HeCTOr cryogenic He target \cite{Hector} with a dedicated mechanical support to hold its cryostat. 

The position of the detectors inside the chamber is determined with a 3D arm which measures the position of holes made in the PCB frame of the detectors. Each detector is measured individually and the target position is measured consistently with a mask set at the target position. The precision of the measurement is 0.2\,mm.

\subsection{The AGATA array}

Details about the AGATA array at GANIL have been published in ref.\cite{agata}. During the MUGAST campaign, the AGATA spectrometer consisted of up to 41 Ge crystals organized in triple clusters and placed in the backward hemisphere of the laboratory frame 18\,cm away from the target. Resulting efficiency, taking into account the presence of the MUGAST detectors, is discussed in section \ref{effAGATA}.

\subsection{VAMOS configuration}

The large acceptance VAMOS++ magnetic spectrometer~\cite{Rej11} was centered at 0\,degrees to detect the forward focused beam-like residues of the reactions. The spectrometer entrance was placed at 73\,cm from the target resulting in an angular acceptance of 4.6 degrees in the current configuration. The focal plane was equipped with $150\times1000$\,mm$^2$ multi-wire parallel plate avalanche counters (MWPPAC), a pair of position sensitive drift chambers (DC) operated at 6\,mbar of isobutane and a stack of segmented ionization chambers operated between 50 and 120\,mbar of CF$_4$. The MWPPAC provided measurement of the time-of-flight relative to the radio-frequency (RF) of the beam. The DC pair allowed measurement of the track of the heavy ions and the segmented ionisation chamber was used to measure energy loss and total energy of the ions. The identification of the ions is obtained based on the time-of-flight, rigidity and energy measurement following the  reconstruction procedure discussed in Ref.~\cite{Rej11}. A dedicated focalized optics is used to ensure a focalisation of the beam at the entrance of the focal plane. Depending on the beam intensities, the focal plane was equipped with a movable finger/shutter, placed upstream from the detection system to block the direct beam from entering the focal plane.

\section{Electronics and data acquisition}

\begin{figure*}[h!]
\begin{center}
\includegraphics[width=\textwidth]{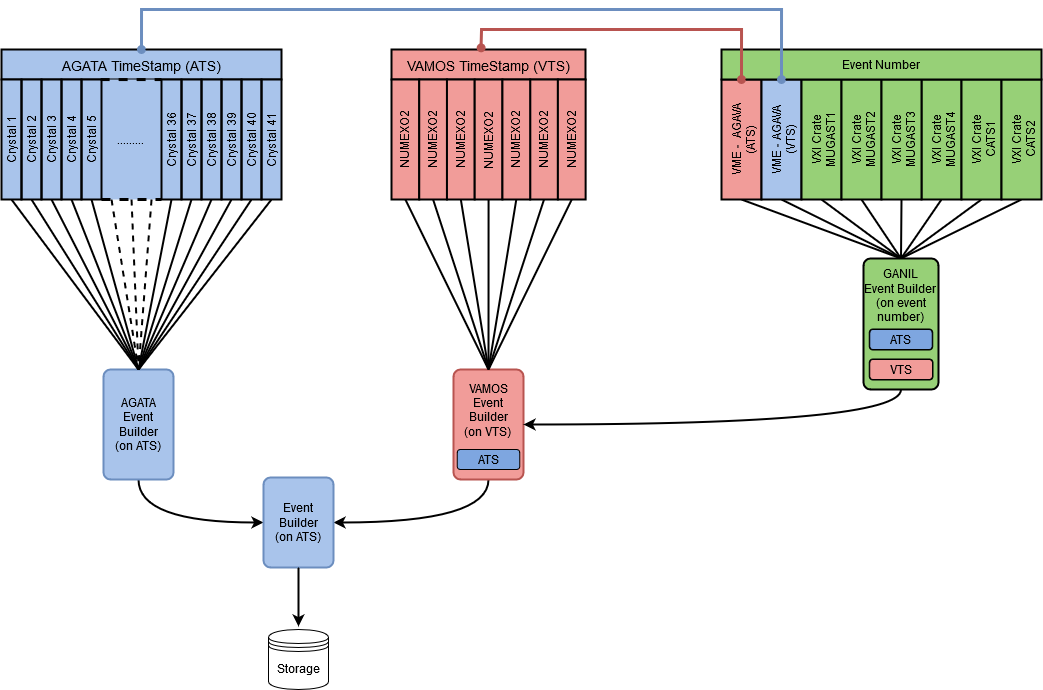}
\caption{Schematic representation of the acquisition and data flow for MUGAST-AGATA-VAMOS. \label{acq}}
\end{center}
\end{figure*}

The MUGAST, AGATA and VAMOS electronics are based on different data acquisition (DAQ) technologies which will be presented in this section before discussing the coupling of the three systems.

\subsection{MUGAST electronics}
The MUGAST array comprises about 3000 strips, all read out by the integrated electronics of MUST2 \cite{must2}, which provide energy and time information for each strip. The front-end boards, called MUFEE, are placed under vacuum at a distance between 20 and 40\,cm from the detectors on a ring surrounding the target (see Fig.\,\ref{design}). Each board hosts 9 ASICs and a pair of MUFEE boards reads 256 strips, i.e. one DSSD. The power consumption is approximatively 15\,W per DSSD and is drained via a water cooled copper heat exchanger sandwiched between the two MUFEE boards.
The ASICs deliver energy, time and leading edge discriminator (LED) information via a multiplexed bus to the MUVI board. MUVI is a single width unit in VXI C standard and ensures the slow control and data coding for 4 telescopes. The full MUGAST electronics comprise 4 MUVI boards in 4 common dead time VXI crates with parallel readout to reduce dead time.

\subsection {AGATA electronics}
The AGATA front-end electronics are based on a full digital system using 14 bit FADC at 100 MHz sampling. The clock and trigger are distributed to each AGATA crystal independently and correlated to the VAMOS and MUGAST sub-system through the AGAVA VME board. Details can be found in Refs.\cite{agata,agava}. In the present configuration, the AGATA electronics were run in trigger-less mode, with no data reduction, since the individual counting rate per detector during the beam time was barely above that of the room background. Pulse-shape analysis of digitized traces from all AGATA channels is performed in real-time and hit positions and energies are stored on disks.
Individual AGATA crystal data are then merged by software with a time coincidence window of 500\,ns using the individual time stamps of each channel. Finally, the coincidence of heavy-ions in VAMOS and light charged particles with AGATA events is performed on the timestamp difference between the AGATA crystals and the AGAVA board with a maximum time window of 2 $\micro$s.

\subsection{VAMOS electronics}
The readout of the detection system of the VAMOS spectrometer was obtained using  NUMEXO2 digitizer modules~\cite{Houarner2021}. Each module consists of 16 channel independent FADC with 100\,MHz sampling. The VAMOS  GTS clock \cite{bellato} is  distributed over each NUMEXO module independently and correlated to the VAMOS and MUGAST sub-system with an AGAVA VME board. In the present configuration, the pre-amplifier outputs of the 10 pads of the ionization chamber and the 4 amplification wires of the drift chambers were digitized and treated in real-time using a trapezoidal filter so that only amplitude and time information are stored upon an external validation of the MUGAST triggered acquisition signal.  The time-of-flight information between the parallel plate detector at the focal plane and the MUGAST array or the RF of the cyclotron or CATS detectors (depending on the configuration) were measured using ORTEC TACs, which signal outputs were digitized using a dedicated NUMEXO2 firmware. The charges of the 4 drift chambers are read out using GASSIPLEX chips and a CAEN V551 sequencer \cite{Rej11} where the charges are multiplexed. The  demultiplexing of the signal is done using NUMEXO2 modules with dedicated firmware.

\subsection{Data acquisition}

The MUGAST-AGATA-VAMOS campaign couples three heterogeneous acquisitions systems: VAMOS and AGATA are based on the GTS system and events are respectively timestamped with the VAMOS TimeStamp (VTS) and AGATA TimeStamp (ATS); the MUGAST acquisition system is based on the GANIL DAQ, coupling 8 VME and VXI crates in common dead time through CENTRUM modules in event number mode \cite{CENTRUM}.   
The MUGAST acquisition system is triggered by an OR of the individual LED signal from the MUGAST telescopes. The VAMOS events are validated by triggered events in MUGAST to achieve data reduction. The AGATA system operates in trigger-less mode. Triggered events in the MUGAST system are timestamped with both ATS and VTS clocks using two AGAVA modules \cite{agava}. A scheme of the data acquisition and data flow is given in Fig.\,\ref{acq}. The correlation of data produced by the systems proceeds in three steps : 
\begin{enumerate} 
\item On the MUGAST subsystem, the events are built on a common dead time event number basis in the GANIL event builder. The ATS and VTS timestamps are embedded in the built event headers.
\item A second event builder, based on the VAMOS TimeStamp (VTS) is used to correlate the VAMOS and MUGAST data. A typical building window of 1 $\micro$s  was used.
\item A third event builder, based on the AGATA TimeStamp (ATS) is used to correlate the  pre-built AGATA \cite{agata} data and the VAMOS+MUGAST data. A typical building windows of 1 $\micro$s was used.
\end{enumerate} 


\section{MUGAST performances}
The MUGAST array provides particle identification and precise energy and angle measurement of the light ejectiles. Some of its performances are discussed in this section.

\subsection{Particle identification}
\begin{figure}[h!]
\begin{center}
\includegraphics[width=0.45\textwidth]{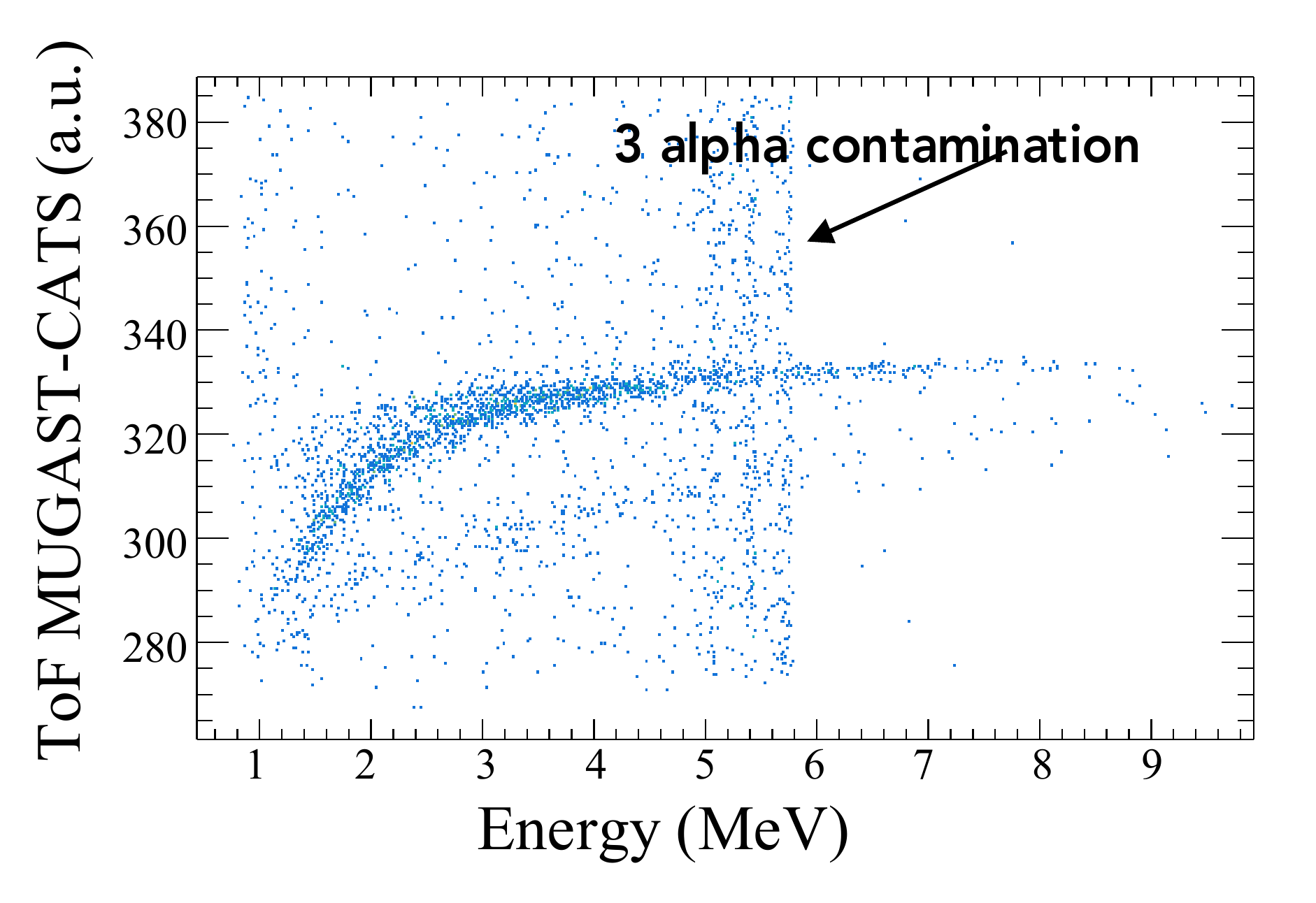}
\caption{ Reverse time-of-flight between MUGAST and the beam tracking detectors CATS as a function of the energy loss in the DSSDs. Note that for reverse ToF, the fastest particles have the longest ToF. A contamination from the triple alpha source is also present.\label{tof}}
\end{center}
\end{figure}

The identification of particles that stop in the first layer of Silicon is usually deduced from time-of-flight (ToF) measurement. In the present case, the ToF is taken between MUGAST and the CATS beam tracking device placed upstream. The typical time resolution is 1.5 ns. An example of particle identification from the time-of-flight versus energy spectrum is shown in Fig.\,\ref{tof} for the $^{16}$O(d,p) reaction measurement. The line corresponding to the protons is clearly seen, with the punchthrough at 9\,MeV for a 500 $\micro$m-thick detector. 

\begin{figure}[h!]
\begin{center}
\includegraphics[width=0.45\textwidth]{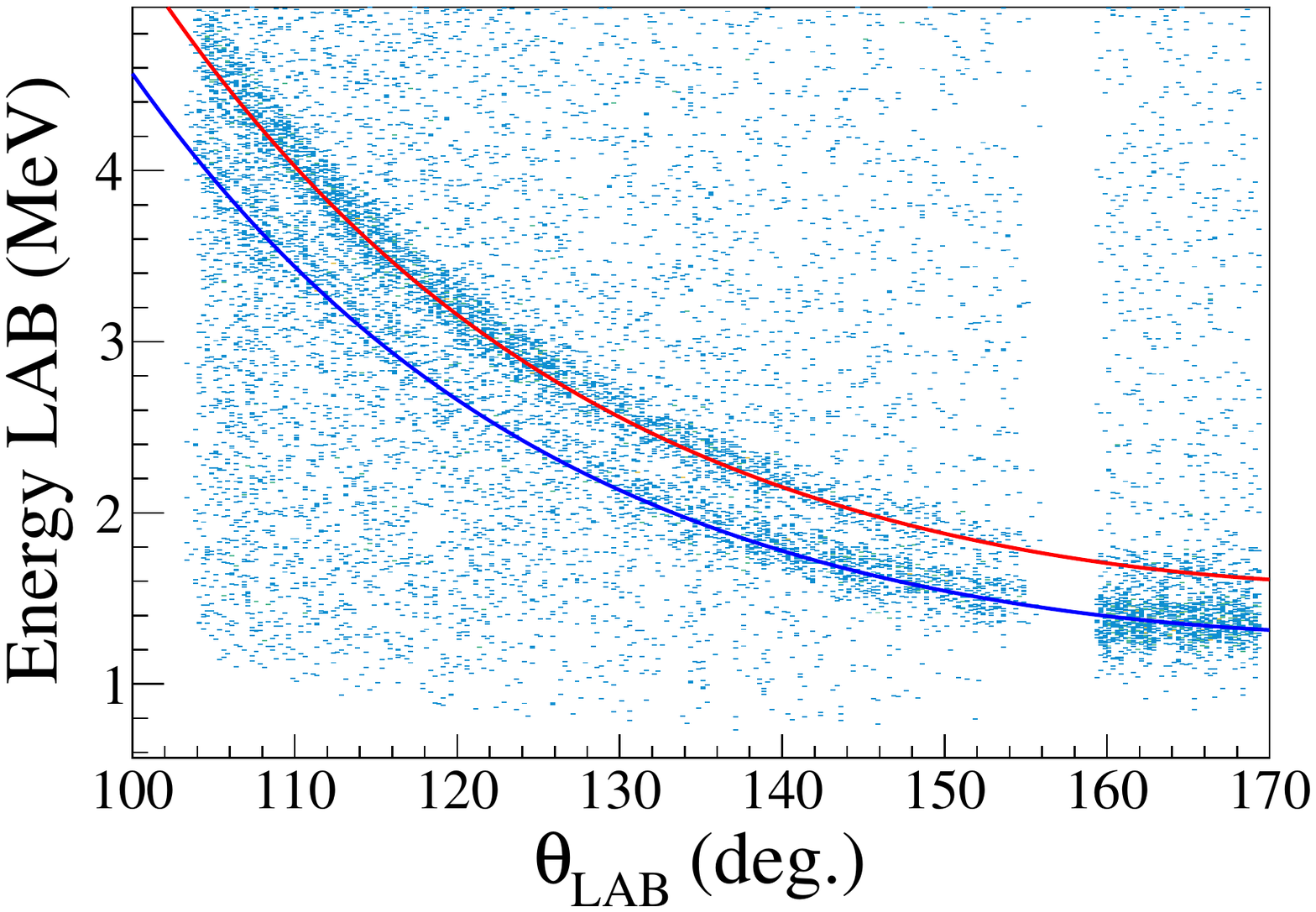}
\caption{ Energy versus angle of the light ejectiles (protons in this case) for the $^{16}$O(d,p) measurement. The red and the blue lines represent the theoretical correlation from two-body kinematics for ground state and the first excited state, respectively. \label{kine}}
\end{center}
\end{figure}

Depending on the beam mass, charge and kinetic energy, the CATS devices cannot always be inserted due to large straggling effects.
When no time reference is available, the particle identification at backward angles relies on the energy-angle kinematic correlation for a two-body reaction, as the background contribution is quite low. From the energy and angle matrix shown in Fig.\,\ref{kine}, the $^{16}$O(d,p) reaction kinematic lines are identified (the solid lines correspond to the theoretical lines). The hole around 155 degrees corresponds to the detector frames. The ground state and the first excited state at 871 keV of $^{17}$O are well separated. The energy resolution for this measurement will be discussed in section \ref{sec_comm}.

During the MUGAST-AGATA-VAMOS campaign, another technique for particle identification was developed, based on the reconstruction of the trajectories in the VAMOS spectrometer as it will be detailed in section \ref{PId_vamos}.

In the forward direction, in addition to ToF identification useful for low energies, particles having enough energy to punch through the first layer (DSSD) are identified from their energy loss in the first layer ($\Delta$E) and their remaining energy measured in the CsI crystals.

\subsection{Excitation energy resolution }
\label{sec_Ex}
The excitation energy of the populated nucleus of interest is deduced by the missing mass method relying on the two-body kinematics and the measurement of the energy and angle of the light ejectile.
For inverse kinematics measurement, the resolution in excitation energy is dominated by target effects and kinematic compression, so that the typical values are several hundreds of\,keV. The thinner and the more homogeneous the target, the better the excitation energy resolution.

\begin{figure}[h!]
\begin{center}
\includegraphics[width=0.45\textwidth]{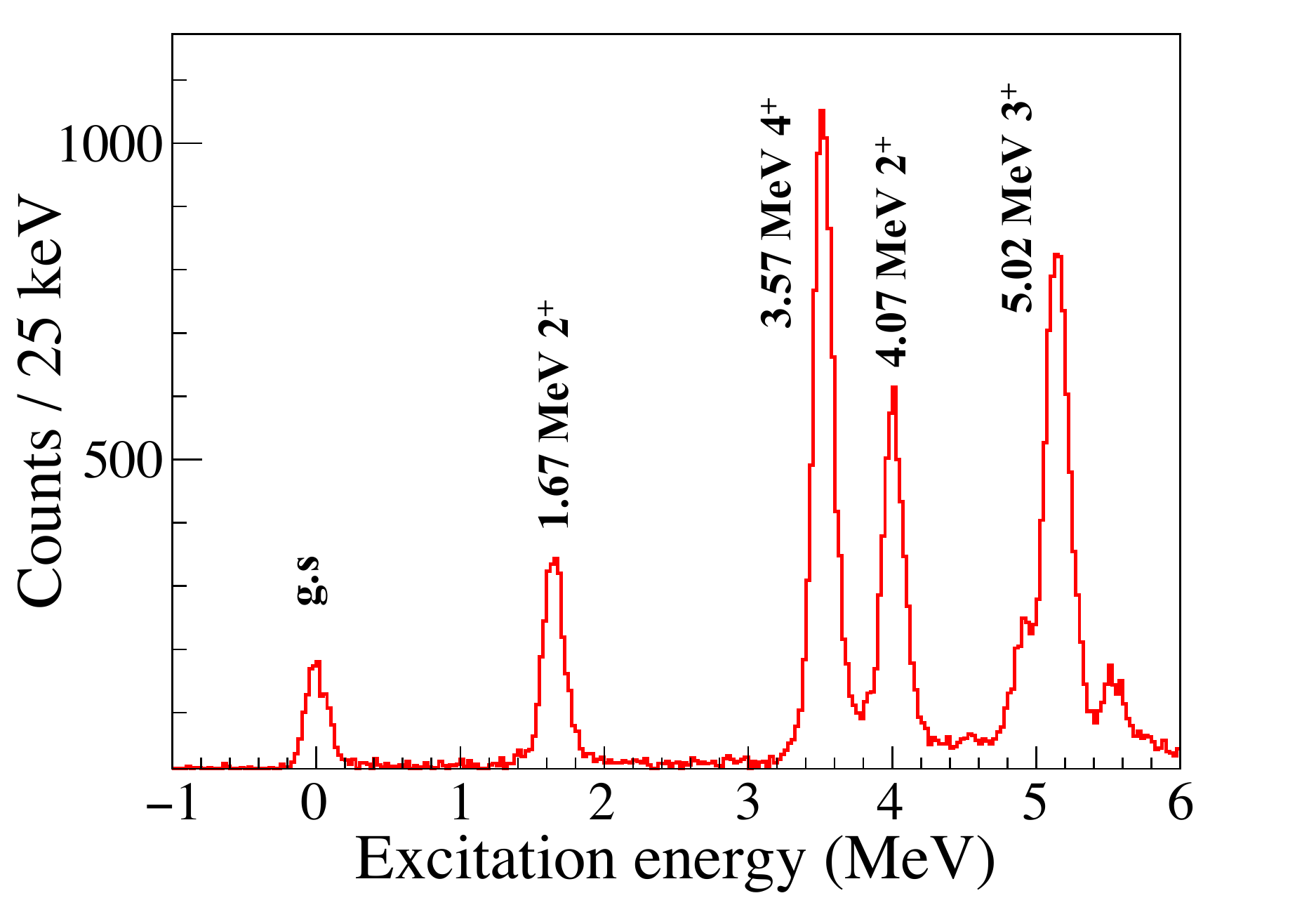}
\caption{ Excitation energy spectrum for $^{19}$O(d,p) reaction with a time gate  between VAMOS PPAC and the RF of the beam to select $^{20}$O. The main peaks observed are labeled.\label{Ex}}
\end{center}
\end{figure}

During the campaign, the thinnest target was used for the measurement of $^{19}$O(d,p$\gamma$)$^{20}$O at 8A\,MeV with a 0.3 mg/cm$^2$ (corresponding to 3 $\micro$m) thick CD$_2$ target. 
The measured excitation energy spectrum is shown on Fig.\,\ref{Ex} ,where the background contribution is removed by gating on the time between the PPAC of VAMOS and the RF of the beam that selects $^{20}$O ejectiles. The experimental excitation energy resolution is 179$\pm$4\,keV for the 2$^+$ excited state of $^{20}$O at 1.67\,MeV.
A similar measurement was performed with HELIOS at 8.06A\,MeV and a slightly thinner target of 0.26 mg/cm$^2$ \cite{Hof12}. An excitation energy resolution of approximatively 175 keV (FWHM) was obtained, very similar to our value. 

\begin{figure}[h!]
\begin{center}
\includegraphics[width=0.45\textwidth]{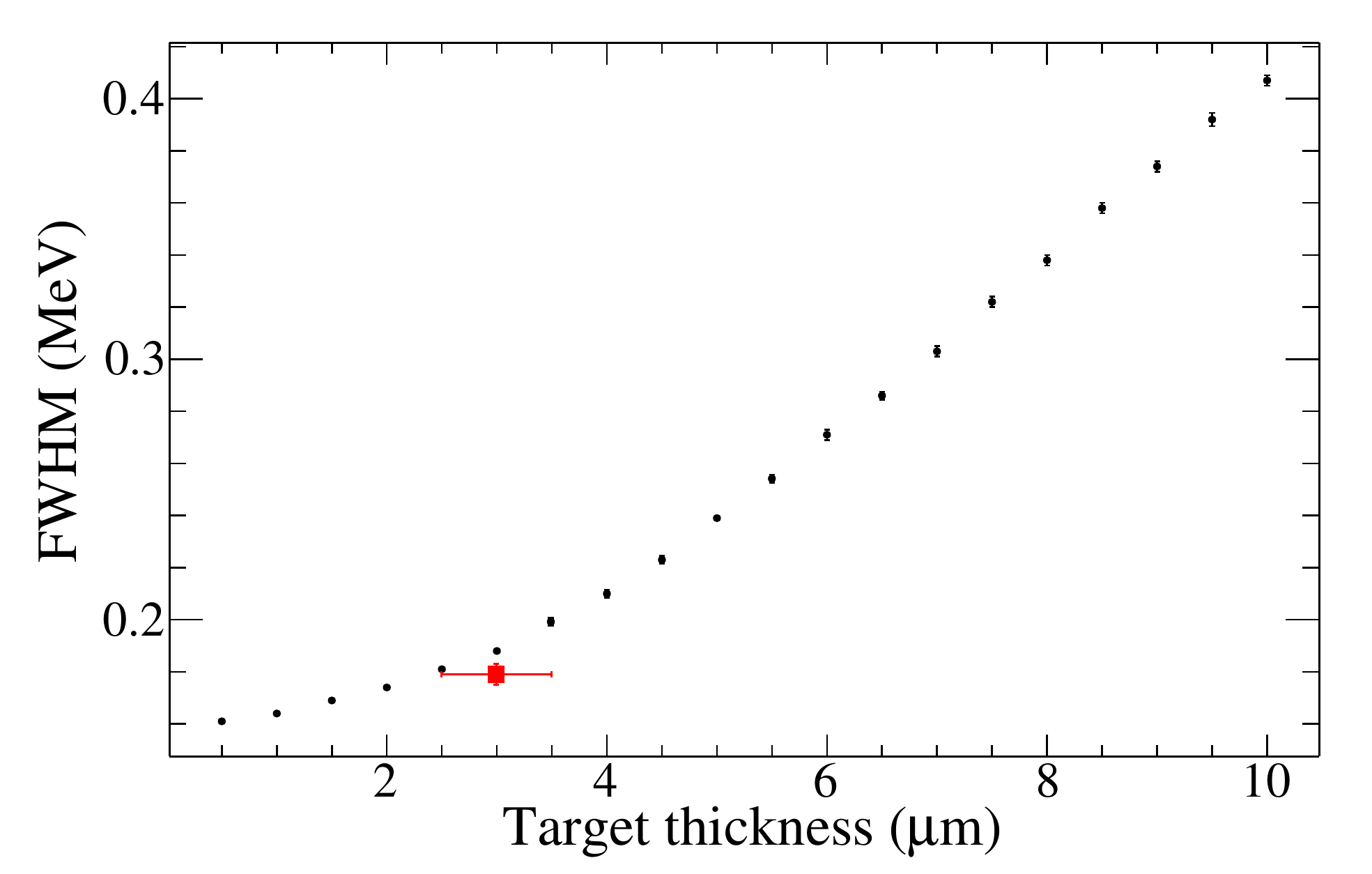}
\includegraphics[width=0.45\textwidth]{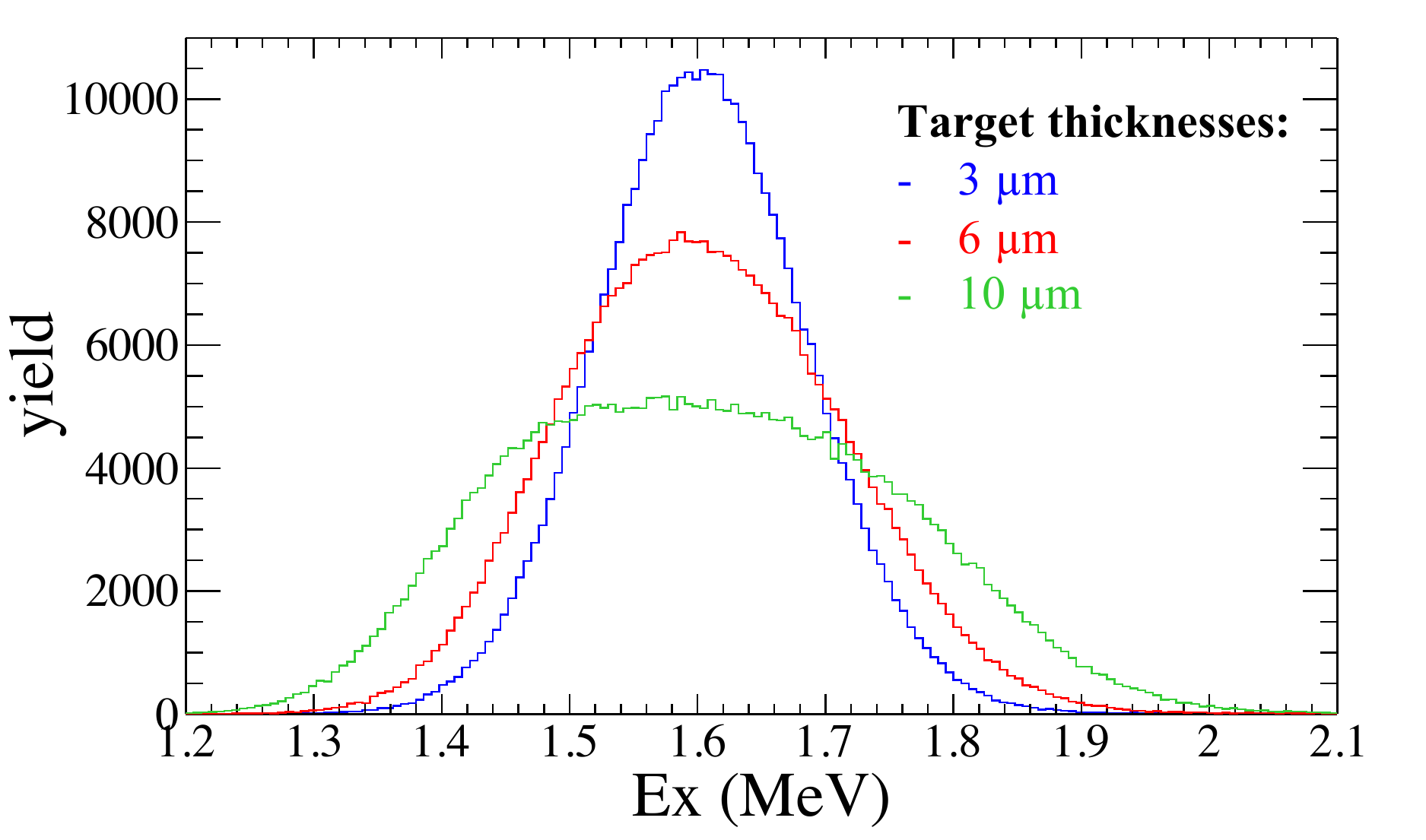}
\caption{(Color online)(Top) Full width half maximum of the excitation energy peak (see text for definition) as a function of the target thickness (in $\micro$m) for the $^{19}$O(d,p)$^{20}$O* reaction, populating the 2$^+$ state at 1.67\,MeV. The red point represents the experimental resolution. (Bottom) Simulated excitation energy spectra for three CD$_2$ target thicknesses: 3 (blue), 6 (red) and 10 (green) $\micro$m for the same reaction. \label{target}}
\end{center}
\end{figure}

In order to get deeper insight into the contributions to the excitation energy resolution, simulations using the \emph{nptool} package are performed in a consistent way with the analysis of our data for the measurement $^{19}$O(d,p$\gamma$)$^{20}$O. The impact of the target thickness on the excitation energy resolution of the 2$^+$ excited state of $^{20}$O is shown on Fig.\,\ref{target} (top). The simulation taking into account the target thickness and the associated straggling in energy and angle, the energy resolution of the detectors and the beam spread on target leads to an excitation energy resolution of 191 keV. The experimental result is shown by the red point for comparison. The uncertainty on the target thickness as given by the manufacturer is quite large 3$\pm$0.5\,$\micro$m and may explain the discrepancy between the simulated and experimental resolutions.
In principle, the target thickness can be accurately determined with the VAMOS spectrometer, taking advantage of the B$\rho$ resolution of 2 per mil. The energy loss of the beam inside the target is precisely determined and the target thickness deduced. But this measurement could not be performed for the $^{19}$O(d,p$\gamma$)$^{20}$O reaction because we were using a gold backing on the target for lifetime measurement purpose. 

Furthermore, the shape of the excitation energy spectrum is also investigated with the simulations in Fig.\,\ref{target} (bottom) for three different target thicknesses 3, 6 and 10 $\micro$m. With increasing target thickness, the Gaussian shape of the excitation energy peak converts into a bell shape. This effect is attributed to the energy loss of the beam in the target and has been previously observed, for example in ref.\,\cite{Bur14, Gir11}. If not properly accounted for, it impacts the extraction of the angular distributions and associated cross-sections. The excitation energy spectrum should, in this case, be fitted with the convolution of a Gaussian and a step function. The full width half maximum (FWHM) plotted in Fig.\,\ref{target} (top) corresponds to the width of the peak evaluated at half the height of the peak by an iterative method, after obtaining the peak height from the fitting procedure.

Another contribution to the excitation energy resolution is the beam spread on target.
With SPIRAL1 beams, the quality of the beam optics, with a size on target of $\Delta$X=1.5 to 1.8\,mm and $\Delta$Y=2.7 to 3.0\,mm, offers the possibility to run without tracking detectors. The simulated excitation energy spectrum for $^{19}$O(d,p)$^{20}$O* reaction assuming $\Delta$X=1.5\,mm and $\Delta$Y=2.7\,mm and a target of 3\,$\micro$m, gives a resolution of 191\,keV without beam tracking and 164\,keV with no beam spread.

In total, the experimental excitation energy resolution is 180\,keV where 40\,keV come from the intrinsic detector resolution and about 30\,keV from the spatial spread of the beam.  The dominant contribution is due to angular and energy straggling of outgoing protons inside the target.


\section{AGATA efficiency with MUGAST}
\label{effAGATA}
One of the major requirements for coupling a Silicon devices with a HPGe array is the transparency to $\gamma$-rays. The impact of the MUGAST array lying in between the target and the AGATA array on the photopeak efficiency is investigated in this section.

\subsection{Photopeak efficiency}
\begin{figure}[h!]
\begin{center}
\includegraphics[width=0.45\textwidth]{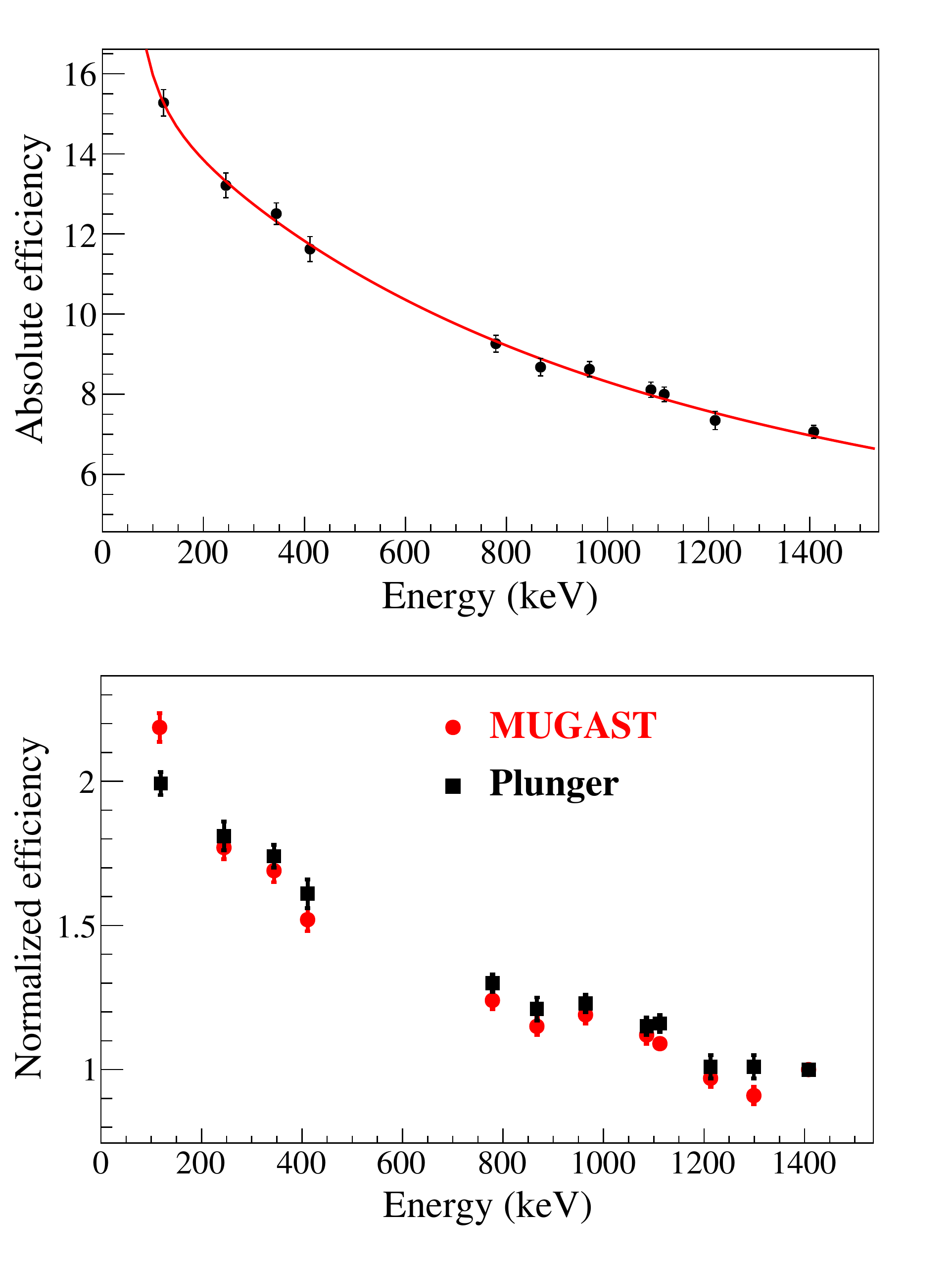}
\caption{ (Top) Absolute efficiency of AGATA with the MUGAST array at 18 cm, after add-back procedure. (Bottom) Comparison of tracked AGATA efficiency between MUGAST set-up (red dots) and the standard chamber with the plunger device (black dots). The efficiency is normalized to the 1.408\,MeV line. \label{effA2}}
\end{center}
\end{figure}

The absolute photo-peak efficiency of AGATA was measured using a standard $^{152}$Eu source placed at the center of the MUGAST array (see Fig.\ref{effA2}(top)). The photo-peak efficiency at 1.4\,MeV  was measured to be 4.9(1)\%  with 41 fully operational crystals placed at 182\,mm from the source 
and 7.1(1)\% after neighboring crystal Add-Back procedure was applied. Nevertheless, the second case results in a worsening of the peak-to-total ratio below 400\,keV as compared to the first case with tracking, as, in fact, is expected from GEANT4 simulations \cite{farnea}. 

\subsection{Transparency to $\gamma$-rays}
The MUGAST array transparency to $\gamma$-rays was carefully evaluated. Fig.\,\ref{effA2} (bottom) presents a comparison between two measurements using a $^{152}$Eu source placed at the target position. The first measurement (black points) was performed with 35 crystals at the nominal distance of 233\,mm using the standard chamber and a plunger device \cite{agata}. The second measurement (red points) was performed with 39 crystals at 182\,mm, placing the source in the MUGAST chamber, including fully equipped silicon detectors. As the number of available AGATA crystals and the target to detectors distances are different between the two measurements, they have been normalized to the photo-peak efficiency at 1.4\,MeV for the $^{152}$Eu source. The comparison shows that the MUGAST chamber and detectors do not induce additional absorption for $\gamma$-ray energies above 200\,keV and is $\sim$10\% more transparent than a plunger set-up at 100\,keV.


\section{VAMOS performances }
The VAMOS magnetic spectrometer provides full identification of the heavy residues from the reaction as well as full rejection of parasitic reactions such as fusion-evaporation when combined with MUGAST identification of light particles.

\begin{figure}[h!]
\begin{center}
\includegraphics[width=0.5\textwidth]{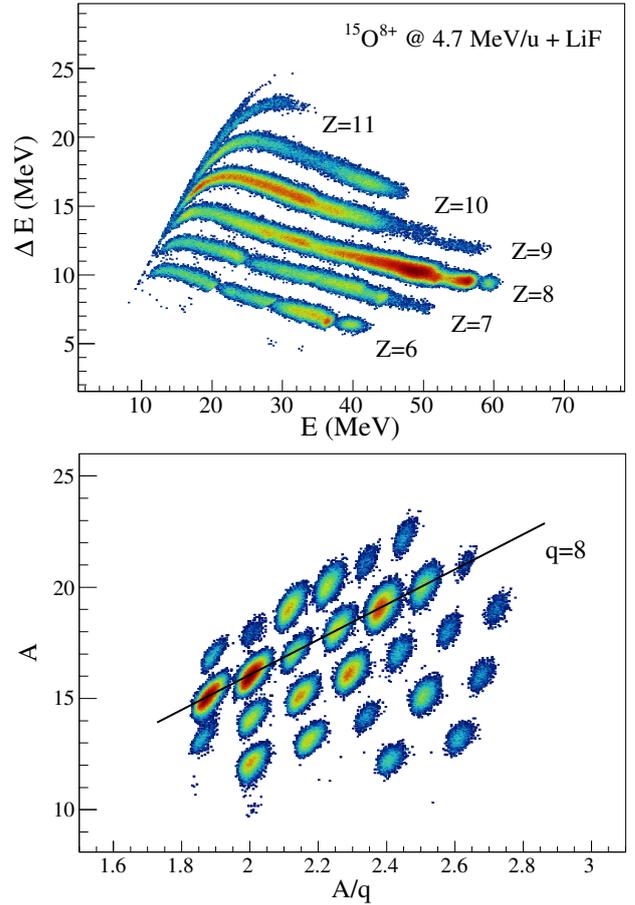}
\caption{Particle identification in VAMOS. (Upper) Identification of the atomic number of heavy residues from $^{15}$O +LiF reactions. (Lower) Identification of mass number and charge states of the residues from $^{15}$O +LiF reactions. The line indicates the charge state q=8. }\label{VAMOS_ID_1}
\end{center}
\end{figure}

The atomic number (Z) is identified based on the correlation between the energy loss ($\Delta$E) in the two first segments of the ionization chamber and the total energy of the ions (E). No overlap between neighboring Zs is observed above the Bragg peak. Fig.\,\ref{VAMOS_ID_1} (upper) shows the Z identification of residues from reactions between the $^{15}$O beam at 4.7A\,MeV and the LiF target.

\begin{figure}[h!]
\begin{center}
\includegraphics[width=0.5\textwidth]{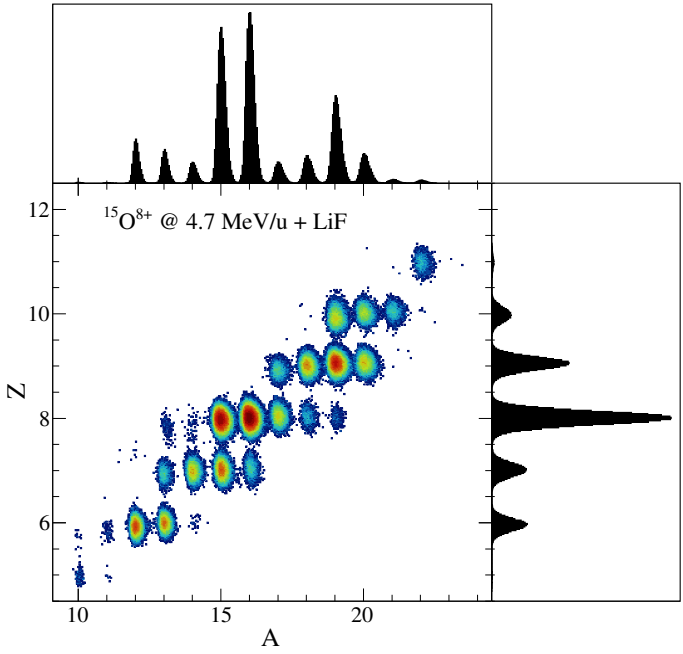}
\caption{Isotopic identification of the heavy residues emitted in $^{15}$O+LiF reactions, detected in VAMOS. Panels in the upper and right edges represent the projection of the mass number and atomic numbers, respectively.}\label{VAMOS_ID_2}
\end{center}
\end{figure} 

 The mass number is obtained based on the combined measurement of the mass, deduced from the energy and velocity of the residues (A = E/u($\gamma$-1)), and the \textit{mass-over-charge} ratio, extracted from their magnetic rigidity (A/q = B$\rho$[Tm]/3.107$\beta\gamma$). The resolution of the mass measurement alone is not good enough to separate neighboring masses, and only the combination of both A and A/q allows for mass separation. Fig.\,\ref{VAMOS_ID_1} (lower) shows the mass identification matrix for the same reactions as in the upper case. The diagonal correlations indicate the different charge states of the emitted ions. 
 
The final isotopic identification of the residues is presented in Fig.~\ref{VAMOS_ID_2}. For the range of masses and energies achieved in this experiment, the resolution in Z results in $\sigma_Z$=1.6\%, in average, and the resolution in A is $\sigma_A$=0.9\%. The resolution in Z is limited by the intrinsic resolution of the ionization chamber and the resolution in A is limited by the time-of-flight between the target and the VAMOS focal plane that, in this case, has a resolution of 2.5 ns.

In the current configuration, the efficiency in the determination of the magnetic rigidity is 90\%, taking into account the drift chamber detection and the reconstruction method efficiencies. The detection efficiency of the ionization chamber is better than 92\% for each section of the detector. 
\section{Performances of the coupled MUGAST, AGATA and VAMOS setup}

The MUGAST-AGATA-VAMOS set-up combines state-of-the-art detectors that were specifically tuned for exclusive measurement of direct reactions. The excitation energy resolution, Doppler correction of $\gamma$-ray lines from particle measurement and developments in VAMOS to accept high beam rates and the advantages of the triple coincidences are discussed in the following sections.

\subsection {Commissioning with $^{16}$O(d,p$\gamma$)$^{17}$O one-nucleon transfer reaction}
\label{sec_comm}
The commissioning of the MUGAST-AGATA-VAMOS campaign was performed with an $^{16}$O beam slowed down to 6A\,MeV and with a reduced intensity of 4$\times$10$^4$ pps impinging on a 1 mg/cm$^2$ CH$_2$ target. The well-studied $^{16}$O(d,p$\gamma$)$^{17}$O reaction \cite{O17} was used as a benchmark. 

\begin{figure}[h!]
\begin{center}
\includegraphics[width=0.45\textwidth]{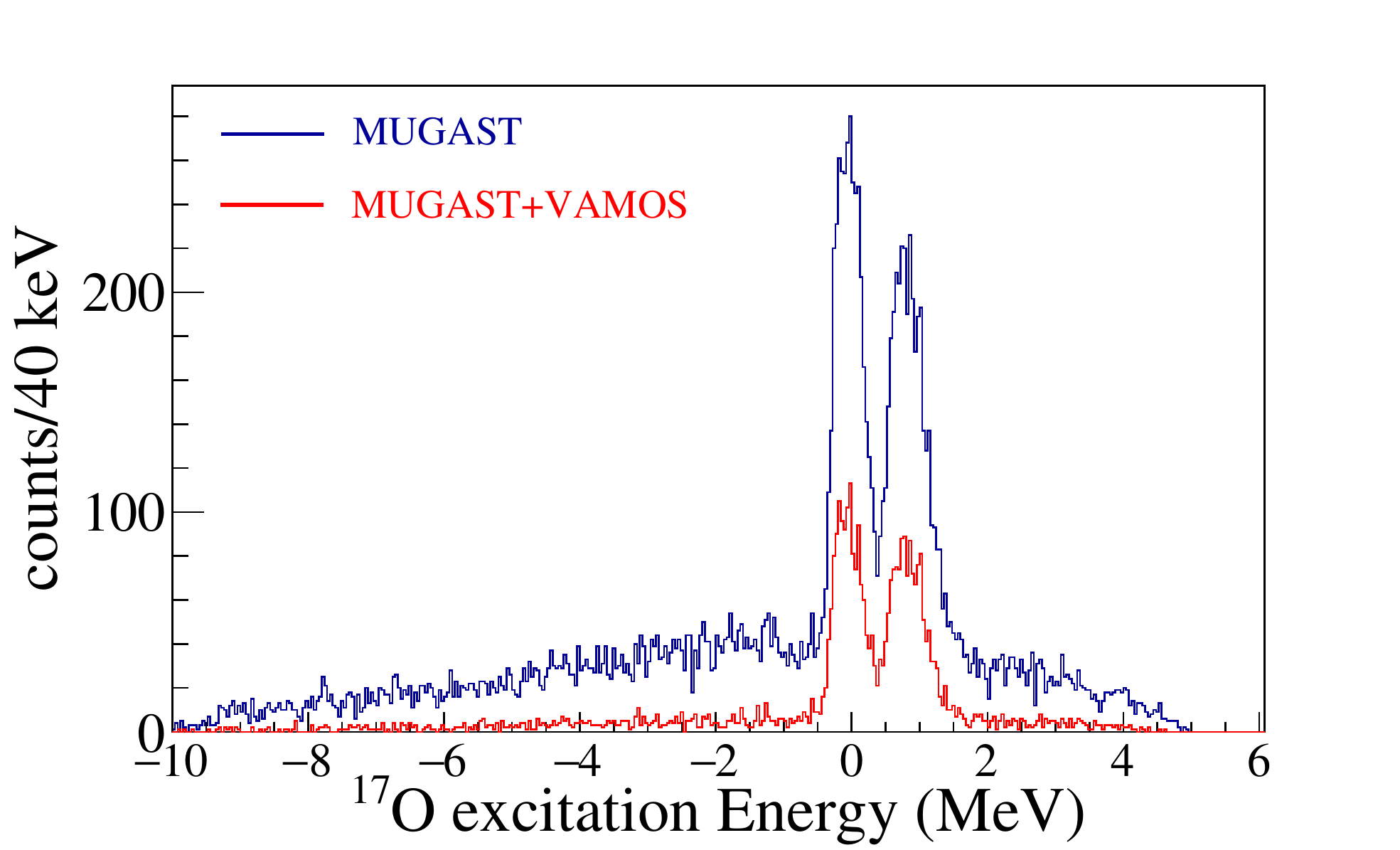}
\caption{ Excitation energy spectra for $^{16}$O(d,p)$^{17}$O obtained from the particle detection alone (MUGAST, blue line) and with an additional condition on the VAMOS ionization chamber (MUGAST+VAMOS, red line). \label{Eexc}}
\end{center}
\end{figure}

The excitation energy spectrum obtained by missing-mass technique is shown in Fig.\,\ref{Eexc}. The background contribution comes mainly from fusion-evaporation reactions and $\alpha$ particles contamination and creates a tail that is clearly visible in the negative energies. A gate on VAMOS suppresses the background at the expense of statistics due to low efficiency during the commissioning experiment.
The excitation energy resolution is 496\,keV and 643\,keV (FWHM) for the g.s. and first excited state, respectively. The main contribution comes from the 1 mg/cm$^2$ thick target effects. In this case, it is sufficient to separate the 871\,keV first excited state from the ground state. 

\begin{figure}[h!]
\begin{center}
\includegraphics[width=0.5\textwidth]{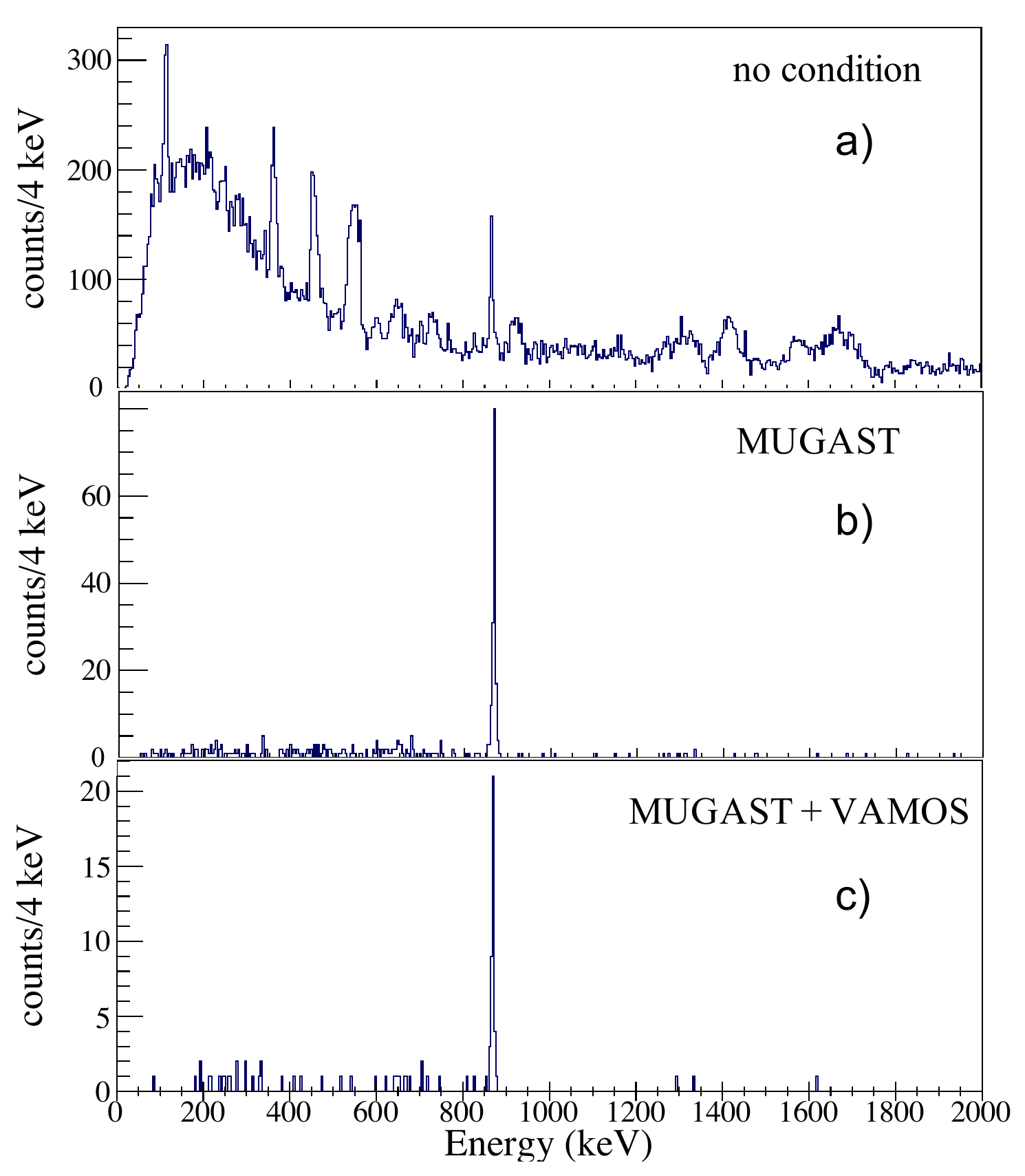}
\caption{ $\gamma$ spectra from AGATA after add-back procedure : a) with a time window between AGATA and MUGAST-VAMOS event and Doppler correction with average $\beta$ of the beam; b) with multiplicity 1 in MUGAST, gating on the excitation energy of Fig.\,\ref{Eexc} on the 871~keV peak, and Doppler correction from the two-body kinematics; c) with an additional condition on the ionization chamber of VAMOS. \label{17Ogam}}
\end{center}
\end{figure}

The resolution can of course be improved by the $\gamma$-ray measurement in coincidence. The Doppler corrected $\gamma$-ray spectra are shown in Fig.\,\ref{17Ogam} (a) without any condition except a time window between AGATA and MUGAST-VAMOS events (b) adding a gate on the excitation energy in MUGAST and finally (c) and adding an extra gate on the ionization chamber of VAMOS. All background contributions are strongly reduced already with a gate on MUGAST while the $\gamma$-ray line at 871\,keV remains very strong.

\begin{figure}[h!]
\begin{center}
\includegraphics[width=0.5\textwidth]{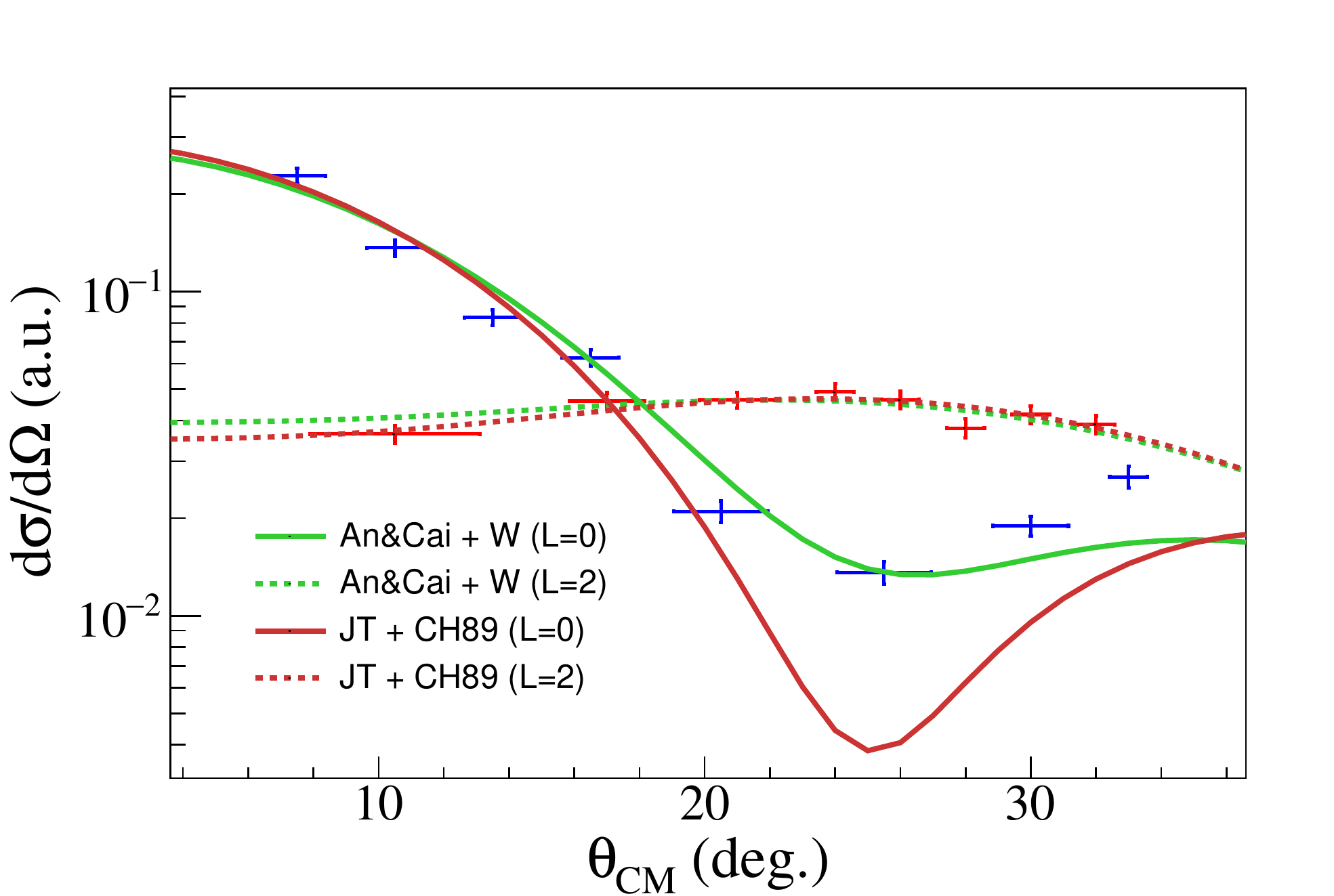}
\caption{ Angular distributions for the g.s. and first excited 1/2$^+$ state in $^{17}$O. The DWBA calculations performed with two different potentials combination are represented in dashed and solid lines for the g.s. and first excited state respectively.\label{Adist}}
\end{center}
\end{figure}

The two angular distributions for transfer to the ground state and first excited state are presented in Fig.\,\ref{Adist}.
They were obtained using a fit of two Gaussians and a background within several, width varying, angular bins. Each bin width was chosen so that the statistics is the same, hence resulting in the same statistical error bars. The efficiency normalization was performed using the \textit{nptool} simulations. 
An absolute normalization was not possible because the beam current was not monitored during the commissioning. This current is normally measured by the CATS beam tracking devices \cite{cats}. When they cannot be used (due to too high current or too large straggling in the device), a model dependent normalisation can be deduced from measurement of the elastic scattering around 90 degrees, however no detector was available in this region during the commissioning run. A relative normalization to the theoretical distributions can still be carried out. 

The transfer reaction is studied with standard post-form DWBA calculations performed with the FRESCO code \cite{FRESCO}. 
For the $^{16}$O-d entrance channel, the parametrization by An\&Cai~\cite{An&Cai} is adopted whereas, for the exit channel and the core-core optical potentials, the global parametrization by Watson et al.~\cite{Watson} is used. The binding potential for the p-n system is taken as the simple Gaussian interaction in \cite{Aust}, which reproduces the deuteron binding energy. For the binding potential of the $^{17}$O nucleus, a Woods-Saxon potential with standard parameters ($r_0=1.25\ \text{fm}, a=0.65\ \text{fm}$) is employed. The depth of the central and spin-orbit terms are adjusted to reproduce the energy of the d$_{5/2}$ ground state and the s$_{1/2}$ excited state in $^{17}$O. In the DWBA calculations, spectroscopic factors equal to unity are fixed for both states, in agreement with previous measurements \cite{Cooper, Tsang}. 

A proper description of the deuteron breakup may play a very important role in the description of the (d,p) transfer process, and could be done using the ADWA formalism. The adiabatic model using the Jonhson-Soper zero-range potential can be used to reproduce this effect \cite{JohnsonSoper}, although it is better suited at high incident energies. The finite-range extension by Jonhson-Tandy (JT) \cite{JohnsonTandy} gives a more reliable description of the breakup process in a broader range. In the present work, calculations using the JT potential in the $^{16}$O-d entrance channel were also performed, and results are found to be similar to those obtained with the global optical potential. 

The theoretical distributions were re-scaled to the experimental ones for comparison purposes. The ratio between g.s. and s$_{1/2}$ scaling factors with the An\&Cai + Watson case is 1.19(6) and is an estimate of the relative spectroscopic factors between the two transfer reactions. The compatibility between the factors is in agreement with the expected single particle behaviour of the residual nucleus and the unitary spectroscopic factors present in literature \cite{Cooper, Tsang}.

\subsection {Triple coincidence measurements: full spectroscopy and background rejection }

\begin{figure}[h!]
\begin{center}
\includegraphics[width=0.5\textwidth]{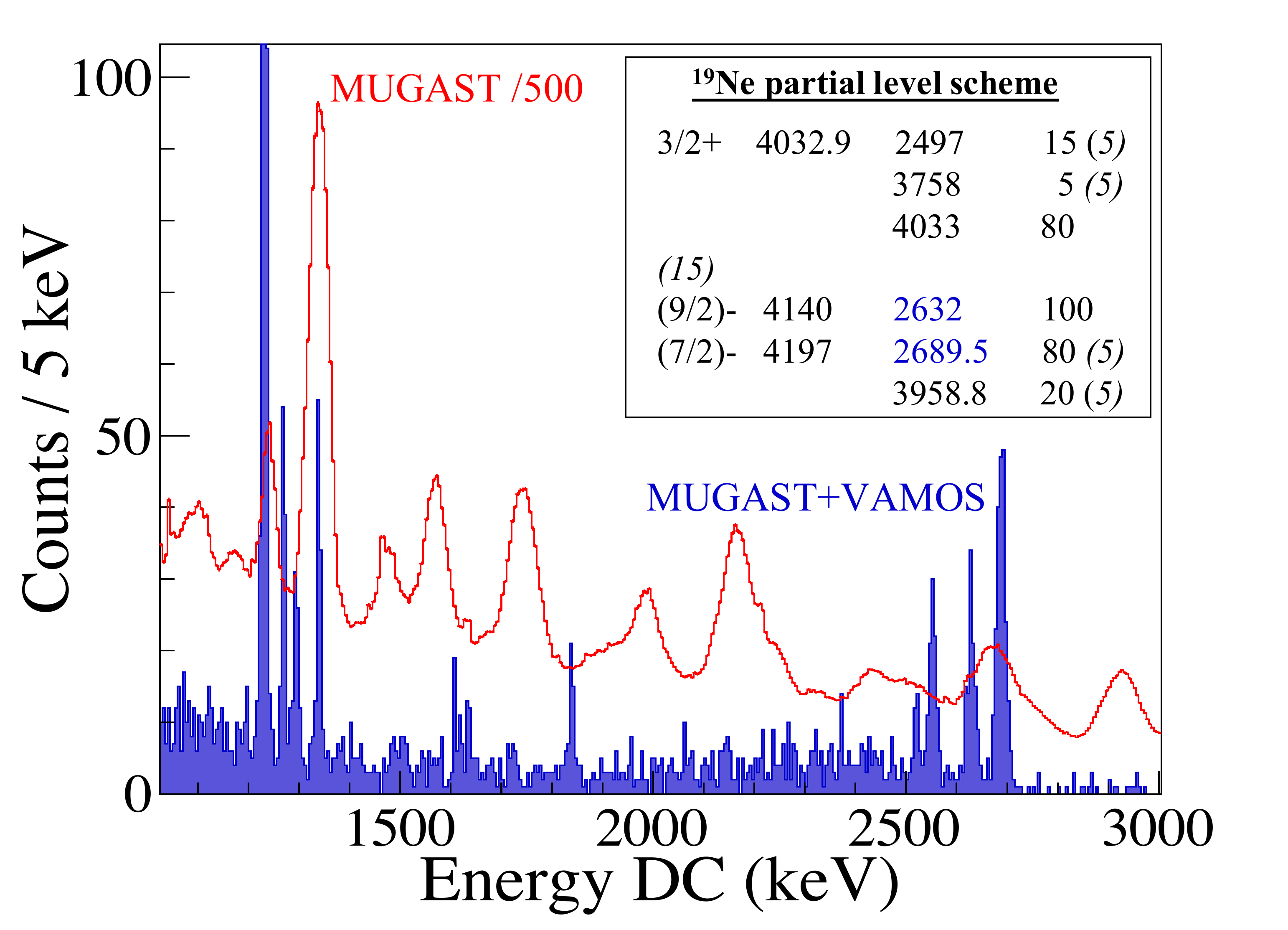}
\caption{ In red, Doppler corrected $\gamma$ spectrum in coincidence with MUGAST (any particle) divided by 500 and in blue the same spectrum gated by MUGAST + $^{19}$Ne ions in VAMOS for the $^{15}$O($^7$Li,t)$^{19}$Ne$^{(*)}$ reaction. A partial list of levels for $^{19}$Ne \cite{ensdf} is given in the inset.\label{Eg}}
\end{center}
\end{figure}

To discuss triple coincidences, we consider the $\alpha$-transfer reaction $^{15}$O($^7$Li,t)$^{19}$Ne$^{(*)}$, of astrophysical interest, that populates a large number of states around 4\,MeV.  For example, the states at 4.140\,MeV and 4.197\,MeV cannot be separated from particle measurement only as they are spaced by 51\,keV. However their de-exciting $\gamma$-ray lines at 2632\,keV and 2689\,keV are well-separated with AGATA (see Fig.\,\ref{Eg}). With triple coincidence measurement, the background contribution to the $\gamma$ spectrum is very low, particularly in the high energy region. 

\begin{figure}[h!]
\begin{center}
\includegraphics[width=0.45\textwidth]{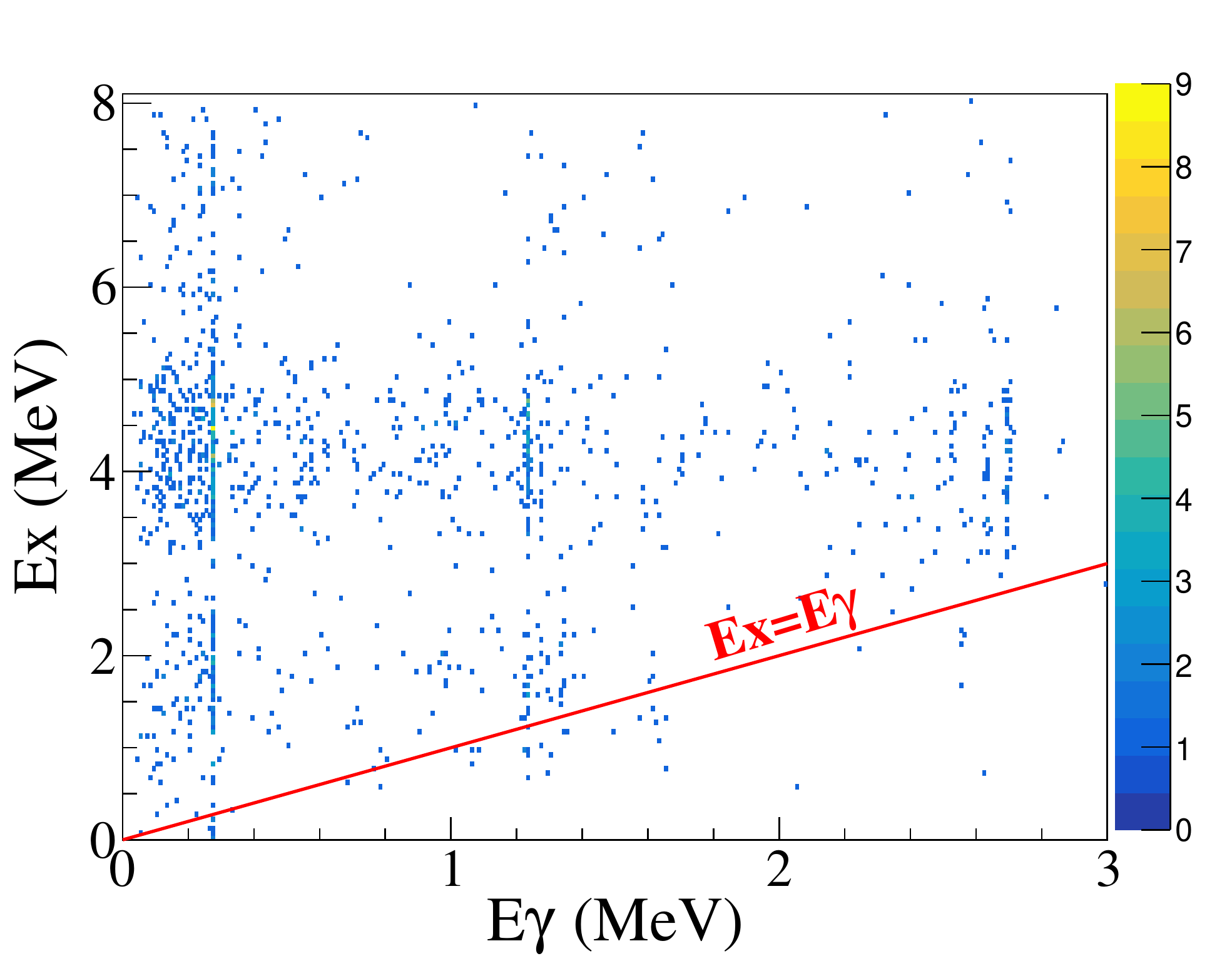}
\caption{ Doppler corrected $\gamma$-ray energy versus the excitation energy for the reaction  $^{15}$O($^7$Li,t)$^{19}$Ne$^{*}$. Counts located on the Ex=E$\gamma$ line correspond to state decaying directly to the ground states of $^{19}$Ne.\label{ExEg}}
\end{center}
\end{figure}

Moreover, the entry point for the $\gamma$ decay can be determined from the particle-$\gamma$ coincidence measurement. Fig.\,\ref{ExEg} shows the Doppler corrected $\gamma$-ray energy (E$_{\gamma}$) plotted against the excitation energy in $^{19}$Ne, as derived from the triton energy and angle. The spectrum is gated by the identified $^{19}$Ne in VAMOS. For example, the 275-keV transition can be seen both through direct population of the 1/2$^-$ state and collecting transition from numerous higher-lying states up to 10\,MeV. This information is useful for determining top feeding of the $\gamma$-rays and can be of interest for lifetime measurements combined with transfer reactions.

\subsection{ Doppler correction of $\gamma$-ray spectra using two-body kinematics}

\begin{figure}[h!]
\begin{center}
\includegraphics[width=0.5\textwidth]{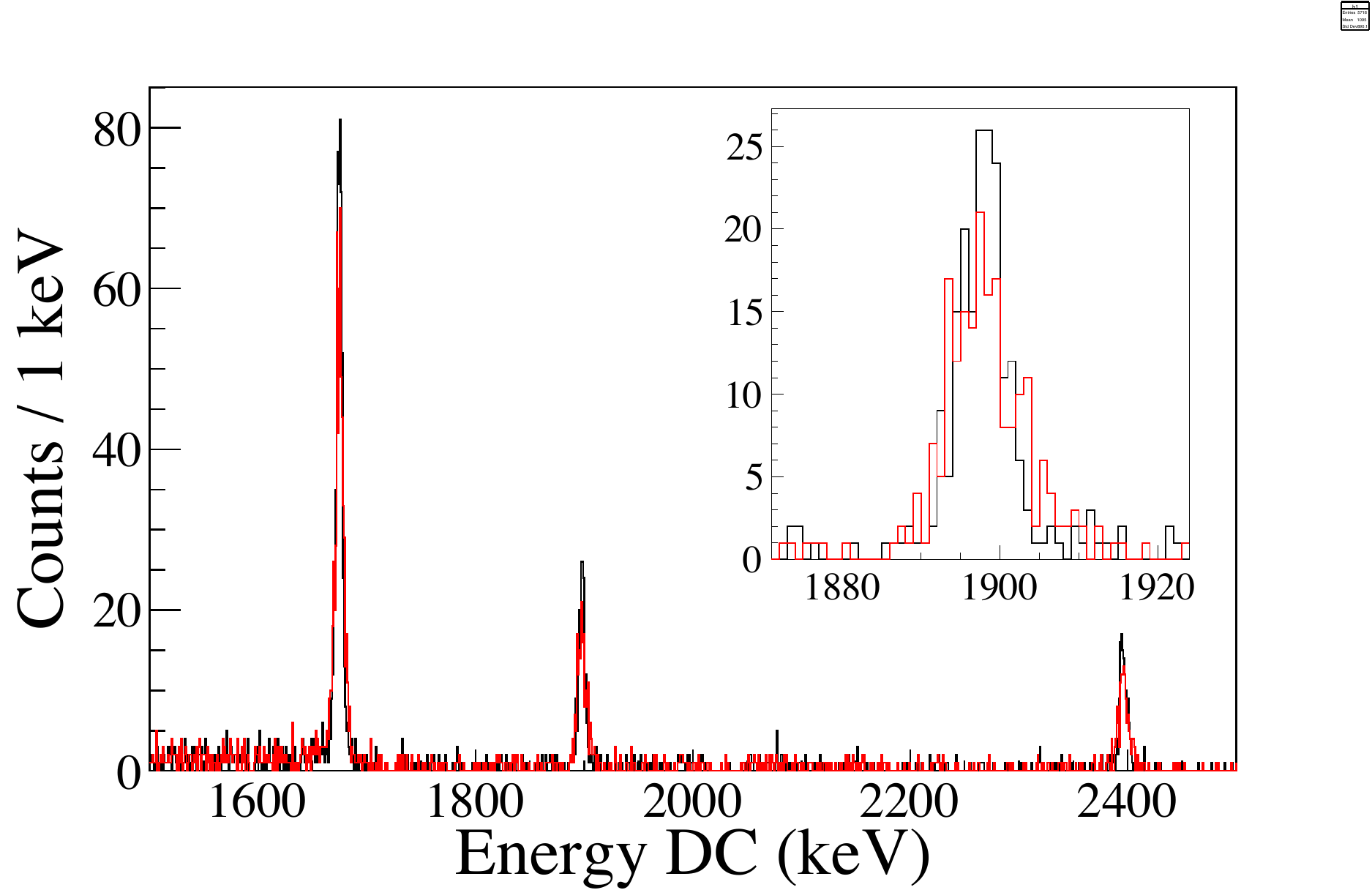}
\caption{ $\gamma$ spectra for $^{19}$O(d,p)$^{20}$O reaction gated on $^{20}$O in VAMOS and Doppler corrected: with the averaged beam energy (red) and on an event-by-event basis with the light particle detected in MUGAST (black). A zoom on the 1.9\,MeV region is shown in the inset. \label{DC}}
\end{center}
\end{figure}

 The Doppler correction of in-flight emitted $\gamma$-rays is usually performed event-by-event using the first hit interaction position determined by the \textit{Orsay Forward Tracking} (OFT) algorithms \cite{oft,agata2} with the positions and energies as provided by the pulse shape analysis algorithms, and the averaged $\beta$ of the beam. With MUGAST, the Doppler correction can also rely on the two-body kinematics by using the momentum of the heavy-ions deduced from the precise position and energy measurement of the light charged particle. A comparison of the two methods is shown in Fig.\,\ref{DC}. A resolution (FWHM) of 7.1\,keV is obtained for the 1.673\,MeV $\gamma$-ray transition from the de-excitation of the 2$^+_1$ state in $^{20}$O, compared to a resolution (FWHM) of 10.2\,keV obtained with the Doppler Correction performed using an average beam velocity of 12.6$\%$.

\subsection{High intensity mode}
\subsubsection{Conditions}
During the MUGAST campaign, radioactive beams of up to 10$^8$\,pps impinged on targets of typical thicknesses of 1\,mg/cm$^2$, with the VAMOS spectrometer set at zero degrees. This high intensity mode was only possible with the masking of the most forward part of the MUST2 detectors and using a movable plate in the VAMOS focal plane to intercept direct beam and non dominant charge states of residues of interest.
The counting rates were 10$^5$ Hz in VAMOS, about 100 Hz for the full MUGAST array and about 100 to 200 Hz per crystal resulting in about 8 kHZ in total in AGATA.

\subsubsection {Particle identification and time-of-flight with VAMOS}
\label{PId_vamos}

One of the difficulties with the high intensity mode is the impossibility to use beam tracking devices for both  normalization of the beam intensity and particle identification by time-of-flight.
For the latter, the time between MUGAST and VAMOS has been investigated by reconstructing the trajectories of heavy-ions in VAMOS. 

Once the residues are identified, the start of the time-of-flight may by traced back from the VAMOS detection. The exact (integer) values of the mass number ($\lfloor$A$\rfloor$) and the ion charge ($\lfloor$q$\rfloor$) allow to translate the magnetic rigidity into velocity:
\begin{equation}
    \beta =\left( \left( \frac{3.107}{B\rho [Tm]}\frac{\lfloor A\rfloor}{\lfloor q\rfloor}\right)^2 +1\right)^{-1/2}.
\end{equation}
This velocity has inherited the resolution of the magnetic rigidity (0.2\%). Hence the time-of-flight between the target and the VAMOS focal plane ($\mbox{ToF}_{VAMOS}$) is accurately determined from this velocity and the path travelled by the residues, also reconstructed in VAMOS with a resolution of 0.2\%. The deduced time-of-flight resolution between the target and VAMOS is $\sim$800\,ps, a factor 4 better resolved than using the cyclotron RF as the start of the time-of-flight measurement. 

The time-of-flight between the target and MUGAST (ToF) may be determined using the previous method, together with the time between MUGAST and VAMOS, $\mbox{ToF} = \mbox{ToF}_{\text{VAMOS}}-\mbox{T}_{\text{MUGAST-VAMOS}}$, with an inherent resolution of $\sim$1 ns.

\begin{figure}[h!]
\begin{center}
\includegraphics[width=0.5\textwidth]{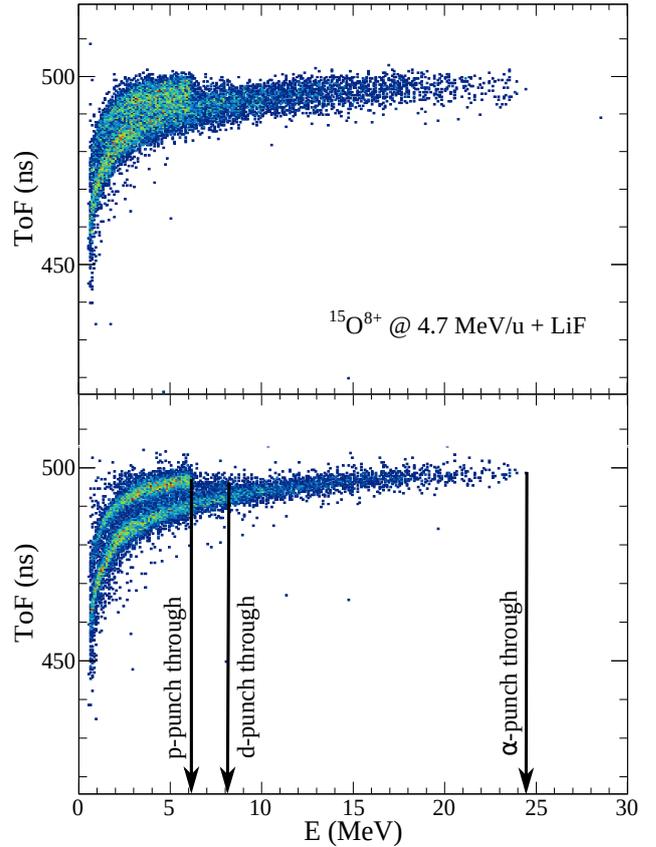}
\caption{ Particle identification by time-of-flight versus Energy. (Upper) The time-of-flight is measured between MUGAST and the RF from the beam. (Lower) The time-of-flight is determined from the time between MUGAST and VAMOS after the reconstruction of the trajectories of the heavy residues.\label{TOF_VAMOS}}
\end{center}
\end{figure}

Figure~\ref{TOF_VAMOS} presents the identification plot in MUGAST based on the relation between the time-of-flight and the energy of the light recoils emitted in reactions between a $^{15}$O beam and a LiF target. 

The upper panel shows the ToF and the energy of the recoils detected in one of the MUST2 detectors placed at forward angles. The ToF is measured as the time difference between the cyclotron RF and the individual signal of each detector strip. The limited resolution of the RF results in a ToF resolution of 2.5 ns and it prevents the separation between different species.

The lower panel shows the same data set, but now, the ToF is determined based on the method presented above. Protons and $\alpha$ particles are well separated. The deuterium line is not visible in the data due to its typical low production in this type of reactions, nevertheless a shadow is present between both proton and $\alpha$ that corresponds to deuterium. The vertical lines indicate the punch-through of proton, deuterium and $\alpha$ particles in the detector. The final ToF resolution is 1.4 ns. This ToF resolution is limited by the experimental conditions: the MUGAST signal used as start of $\mbox{T}_{\text{MUGAST-VAMOS}}$ suffered from electronic jitters that imposed $\sim$1 ns time resolution to the signal.

\section{Conclusion}

The MUGAST-AGATA-VAMOS set-up combines simultaneous light particle, $\gamma$-ray and heavy recoil detection with state-of-the-art devices and allows for triple coincidence measurements of direct reactions with radioactive beams.
The system was commissioned with the $^{16}$O(d,p$\gamma$)$^{17}$O$^*$ reaction for which the spectroscopy of $^{17}$O was performed.
Several stripping reactions measurements (d,p$\gamma$), ($^{7}$Li,t$\gamma$) performed during the campaign highlight the specificity of the MUGAST-AGATA-VAMOS set-up. In particular, it has been demonstrated that the detection of light ejectiles with good accuracy in energy and angle improves the Doppler correction of the $\gamma$-ray energies as compared with the method using the average velocity of the beam. 
The experimental setup shows excellent background rejection in the particle and $\gamma$-ray spectra thanks to the triple coincidence measurement. It also provides the determination of the entry point for $\gamma$-ray decay from the transfer reaction measurement. This information is crucial when applying Doppler Shift Attenuation Method (DSAM) for lifetime measurements.
Moreover, a new method for particle identification based on the time-of-flight between MUGAST and VAMOS by reconstructing trajectories in VAMOS has been proposed. The time resolution obtained by this technique (1.4 ns) allows for particle identification without using beam tracking devices. 
The set-up has also proven to be able to run with high intensity beams (up to 10$^8$ pps), after masking part of the forward MUST2 detectors and using a movable plate in the focal plane of VAMOS. 
The MUGAST-AGATA-VAMOS campaign paves the way for direct reactions measurement with the new generation of post-accelerated ISOL facilities like SPES \cite{spes} and HIE-ISOLDE \cite{isolde} and the slowed down fragmentation beams like OEDO \cite{oedo} in RIKEN or the low energy branch of FAIR \cite{fair} using the GRIT-AGATA \cite{grit} in combination with high performance zero-degree detection.


\section{Acknowledgments}
The authors would like to thank the GANIL staff for their continuous help.
This project has received funding from the European Union’s Horizon 2020 research and innovation programme under grant agreement No 654002.
A. Matta and W.N. Catford gratefully acknowledge the support of the STFC grant ST/J000051/1, ST/L005743/1, and ST/N002636/1. 
C. Diget acknowledges the support of the STFC grant ST/P003885/1.
B. Fernández-Domínguez acknowledges the support of the  Xunta de Galicia ED431B 2018/15 and Mineco  PGC2018-096717-B-C22 grants.

\end{document}